\documentclass[a4paper,11pt]{article}
\usepackage{jheppub} % for details on the use of the package, please see the JINST-author-manual
\usepackage{slashed}
\usepackage{placeins}
\title{
TeV-scale leptogenesis in a parity symmetric neutrino mass model
}

\author[a,b]{Keisuke Harigaya}
\author[a]{and Gio Leone}

\affiliation[a]{%
Department of Physics, Enrico Fermi Institute, Leinweber Institute for Theoretical Physics, and Kavli Institute for Cosmological Physics, University of Chicago, Chicago, IL 60637, USA}
\affiliation[b]{
Kavli Institute for the Physics and Mathematics of the Universe (WPI), The University of Tokyo Institutes for Advanced Study, The University of Tokyo, Kashiwa, Chiba 277-8583, Japan
}
\emailAdd{kharigaya@uchicago.edu}
\emailAdd{grleone@uchicago.edu}

\abstract{
We study leptogenesis via out-of-equilibrium decays of right-handed neutrinos in a parity symmetric extension of the Standard Model with gauge group $SU(3)_c \times SU(2)_L \times SU(2)_R \times U(1)_X$, spontaneously broken to the Standard Model gauge group at a scale $v_R$. We focus on the minimal Higgs realization, in which the strong CP problem is resolved without invoking additional symmetries. In this framework, lepton number is violated through Yukawa interactions, while neutrino masses arise only at higher loop order, a feature that permits the CP-violating quantum corrections to right-handed neutrino decays to be parametrically large. For thermal production of right-handed neutrinos, we find that successful leptogenesis requires $v_R \gtrsim 6\times 10^{12}$ GeV. Non-thermal production, however, relaxes this bound dramatically, requiring only $v_R \gtrsim 10$ TeV, a scale accessible to collider searches for new particles and searches for rare processes.
}


\begin{document}
\maketitle

%\tableofcontents

\section{\label{sec:intro}Introduction}

The observed baryon asymmetry of the Universe remains one of the most compelling open problems in fundamental physics. Its abundance, $n_B/n_\gamma \simeq 6\times 10^{-10}$, is precisely measured through the temperature and polarization power spectra of the Cosmic Microwave Background~\cite{Planck:2018vyg}. This motivates assessing particle-physics models of the early Universe by their ability to reproduce this asymmetry. Starting from an initial state with negligible net baryon number, baryogenesis requires the three Sakharov conditions: (1) baryon number violation, (2) charge (C) and charge-parity (CP) violation, and (3) departure from thermal equilibrium~\cite{Sakharov:1967dj}. In leptogenesis, a net lepton asymmetry is generated by out-of-equilibrium decays of right-handed neutrinos and subsequently converted into a baryon asymmetry via electroweak sphaleron processes~\cite{Fukugita:1986hr,Asaka:1999yd,Buchmuller:2004nz, Buchmuller:2005eh, Davidson_2008}. In this work, we study leptogenesis in a parity symmetric extension of the Standard Model (SM).

Parity symmetric extensions of the SM are well-motivated. By forbidding the parity-odd QCD $\Theta$ term at high energies, they provide an elegant solution to the strong CP problem~\cite{beg_strong_1978,mohapatra1980,Babu:1988mw,babu_solution_1990,Hall:2018let,Hisano:2023izx}. The gauge group is $SU(3)_c \times SU(2)_L \times SU(2)_R \times U(1)_X$. In the minimal Higgs realization, $SU(2)_R \times U(1)_X$ is spontaneously broken to $U(1)_Y$ by the condensation of an $SU(2)_R$ doublet Higgs $H_R$, the parity partner of the SM Higgs $H_L$~\cite{Babu:1988mw,babu_solution_1990}. This model avoids the hierarchy problem between the $SU(2)_R$ symmetry-breaking scale and the fundamental scale~\cite{Hall:2018let, Dunsky:2019api}, while solving the strong CP problem without invoking any symmetry beyond parity itself.

In the minimal Higgs model, the observed nonzero neutrino masses can be accommodated by introducing gauge-singlet fermions $S$ that acquire Dirac masses with the right-handed neutrinos through $SU(2)_R$ symmetry breaking~\cite{Hall:2023vjb}; see~\cite{Babu:1988yq,Babu:2022ikf,Babu:2023dzz} for alternative neutrino mass models. We refer to the resulting Dirac state also as $S$. In this setup, lepton number is violated by the Yukawa coupling of $S$, but in the absence of a Majorana mass for $S$, SM neutrino masses arise only at five-loop order~\cite{Harigaya:2025zru}. This suppression allows the Yukawa coupling to be large while keeping the SM neutrino masses small, which is crucial for enhancing the CP asymmetry in $S$ decays. The out-of-equilibrium decay of $S$, with CP violation sourced by quantum corrections from these Yukawa couplings, then generates a $B-L$ asymmetry. For related work in which $S$ carries large Majorana masses, see~\cite{Dunsky:2020dhn,Carrasco-Martinez:2023nit,Carrasco-Martinez:2025zus,Babu:2025vjm}; for baryogenesis via a first-order $SU(2)_R$ phase transition, see~\cite{Harigaya:2022wzt}.

We focus on the epoch after the $SU(2)_R$ phase transition but before the SM electroweak phase transition. Following $SU(2)_R$ breaking, $S$ decays through two channels: $S \to \bar{e}\,W_R^{-(*)}$, mediated by the $SU(2)_R$ gauge interaction, and $S \to \ell\,H_L$, mediated by the Yukawa interaction of $S$. The coexistence of these two decay modes violates $B-L$, and CP is generically violated by the Yukawa couplings, producing a net lepton asymmetry that sphaleron processes subsequently convert into a baryon asymmetry. Notably, in contrast to conventional leptogenesis scenarios driven by the decays of heavy Majorana right-handed neutrinos, the charged-lepton sector plays a central role here.

This work is organized as follows. In Section~\ref{sec:model}, we introduce a parity symmetric neutrino mass Model, detail the field content and interactions, and discuss how the singlet and charged-lepton sectors furnish the necessary ingredients for leptogenesis. In Section~\ref{sec:CPViolDyn}, we analyze the CP asymmetry and its relevant flavor structures. In Section~\ref{subsec:therm}, we study thermal production of right-handed neutrinos and derive an analytic upper bound on the baryon asymmetry, yielding a lower bound $v_R \gtrsim 6\times 10^{12}$\,GeV on the $SU(2)_R$ breaking scale. In Section~\ref{subsec:nontherm}, we turn to non-thermal production via inflaton decay, and, after imposing washout and reheating constraints, map out the viable low-scale parameter space, where $v_R \gtrsim 10^{4}$\,GeV. We compare this window against current constraints on right-handed gauge bosons from the LHC and the projected sensitivity at the HL-LHC, a muon collider, and a same-sign muon collider ($\mu$TRISTAN).

\section{\label{sec:model}Parity symmetric model} 
We assume the gauge group
\begin{equation}
    SU(3)_c \times SU(2)_L \times SU(2)_R \times U(1)_X
\end{equation}
with two scalar fields $H_L$ and $H_R$ transforming as $({\bf 1},{\bf 2},{\bf 1},1/2)$ and $({\bf 1},{\bf 1},{\bf 2},-1/2)$ respectively. By imposing discrete parity symmetry, which exchanges left-handed and right-handed fermions, the parity-odd QCD $\Theta$-term is forbidden. We recover the Standard Model gauge group via spontaneous symmetry breaking of $SU(2)_R \times U(1)_X$ into hypercharge $U(1)_Y$ with $H_R$ getting a vacuum expectation value of $v_R$. Three gauge bosons $W_R^\pm$ and $Z'$ obtain masses by the Higgs mechanism.
After the $SU(2)_R \times U(1)_X$ breaking, which also breaks the parity symmetry, the QCD $\Theta$-term is radiatively generated, but the correction is below the experimental bound~\cite{Hall:2018let,Hisano:2023izx}.

$H_L$ is the SM Higgs.
As temperature decreases, $SU(2)_L \times U(1)_Y$ is spontaneously broken into $U(1)_{\mathrm{EM}}$, with $H_L$ getting a vacuum expectation value of $v_L < v_R$. We use the $v_L \simeq 174\,$GeV normalization.
There exist several ways to achieve $v_L \ll v_R$~\cite{Babu:1988mw,Blinov:2016kte,Hall:2018let,Baldwin:2025oqt}.

This setup has several phenomenological advantages over another class of models where $SU(2)_L \times U(1)_Y$ is broken by the vacuum expectation values of $SU(2)_L\times SU(2)_R$ doublets. First, since the phases of the Higgses are gauge degrees of freedom, the Higgs vacuum expectation values have no physical CP phases and hence do not introduce a strong CP phase. Second, the intermediate scale $v_R$ does not necessarily introduce a hierarchy problem beyond that of the electroweak scale. In fact, because of the symmetry $H_L \leftrightarrow H_R$, fine-tuning the mass of $H_R$ in comparison with a fundamental scale $\Lambda$ simultaneously fine-tunes the mass of $H_R$ down to $v_R$, so the total fine-tuning may be as small as $(v_R^2/\Lambda^2)\times (v_L^2/v_R^2)= v_L^2/\Lambda^2$, which is the same as the electroweak fine-tuning~\cite{Hall:2018let, Dunsky:2019api}.

\subsection{A radiative neutrino mass model }

The left-handed neutrino masses can be explained by introducing at least two gauge-singlet fermions $S_i$. The mass and interaction terms of $S$ are
\begin{equation}\label{RadL}
    \mathcal{L} \supset - S_i \left(\lambda^{S\,*}_{ia}H_L \ell_a + \lambda^S_{ia}H_R \bar{\ell}_a\right)- \frac{1}{2}M^{\mathrm{Maj}}_{ij}S_i S_j+ \mathrm{h.c.},
\end{equation}
where $\lambda^S$ is the Yukawa matrix and $M^{\mathrm{Maj}}$ is the Majorana mass matrix of $S$. Here $i,j,k$ label singlet generations and $a,b,c$ label doublet generations. The full particle spectrum with gauge charges is shown in Table \ref{tab:LRcharges}.  When $M^{\mathrm{Maj}}$ is nonzero, one-loop corrections by $S$ generate left-handed neutrino masses $m_\nu$~\cite{Hall:2023vjb}. 
The coupling $\lambda^S$ may be as large as $\mathcal O(1)$; the observed neutrino mass can be explained by taking sufficiently small $M^{\mathrm{Maj}}$.%
\footnote{
Even if $M^{\rm Maj}$ is small at tree level, it is non-multiplicatively generated by four-loop corrections, but the resultant correction to the neutrino mass can be small enough even for $\lambda^S=\mathcal O(1)$ and $v_R= O(10)$ TeV~\cite{Harigaya:2025zru}.
}

The case with $M^{\mathrm{Maj}} > \lambda^{S} v_R$ is studied in~\cite{Dunsky:2020dhn,Carrasco-Martinez:2023nit,Carrasco-Martinez:2025zus,Babu:2025vjm}.
In this paper, we instead consider $M^{\mathrm{Maj}} \ll \lambda^{S} v_R$, for which $m_\nu \propto M^{\mathrm{Maj}}$.
As we will see, leptogenesis is successful even for $v_R$ as low as $10$ TeV.
The smallness of $M^{\mathrm{Maj}}$ may be explained by a spontaneously broken gauge symmetry~\cite{Harigaya:2025zru}.  Because $M^{\mathrm{Maj}}$ is small, we may neglect $M^{\mathrm{Maj}}$ in the interactions around the $SU(2)_R$ breaking scale. Even without the Majorana mass term, $B-L$ is violated;
$\ell$ and $\bar{\ell}$ have $B-L = -1$ and $1$, respectively, so the coexistence of the two Yukawa couplings of $S$ in Eq.~\eqref{RadL} violates $B-L$.

\begin{table}[ht]
\centering
\setlength{\tabcolsep}{6pt}
\renewcommand{\arraystretch}{1.6}
\begin{tabular}{|c|c|c|c|c|c|c|c|c|c|c|c|c|c|} \hline
                      & $H_L$          & $H_R$  & $S_i$       & $q_i$         & $\bar{q}_i$     & $\ell_i$        & $\bar{\ell}_i$ & $U_i$         & $\bar{U}_i$     & $D_i$          & $\bar{D}_i$     & $E_i$   & $\bar{E}_i$ \\ \hline
            $SU(3)_c$ & {\bf 1}        & {\bf 1}  & {\bf 1}      & {\bf 3}       & ${\bf \bar{3}}$ & {\bf 1}         & {\bf 1}        & {\bf 3}       & ${\bf \bar{3}}$ & {\bf 3}        & ${\bf \bar{3}}$ & {\bf 1} & {\bf 1}     \\
            $SU(2)_L$ & {\bf 2}        & {\bf 1}  & {\bf 1}       & {\bf 2}       & {\bf 1}         & {\bf 2}         & {\bf 1}        & {\bf 1}       & {\bf 1}         & {\bf 1}        & {\bf 1}         & {\bf 1} & {\bf 1}     \\
            $SU(2)_R$ & {\bf 1}        & {\bf 2}   & {\bf 1}      & {\bf 1}       & {\bf 2}         & {\bf 1}         & {\bf 2}        & {\bf 1}       & {\bf 1}         & {\bf 1}        & {\bf 1}         & {\bf 1} & {\bf 1}     \\
            $U(1)_X$  & $\frac{1}{2}$ & $-\frac{1}{2}$ & $0$ & $\frac{1}{6}$ & $-\frac{1}{6}$  & $- \frac{1}{2}$ & $\frac{1}{2}$  & $\frac{2}{3}$ & $-\frac{2}{3}$  & $-\frac{1}{3}$ & $\frac{1}{3}$   & $-1$    & $1$         \\ \hline
        \end{tabular}
\caption{The particle spectrum for a parity symmetric neutrino mass model with gauge representations and charges.}
\label{tab:LRcharges}
\end{table}

After the $SU(2)_R$ symmetry breaking, from Eq.~\eqref{RadL}, Dirac mass terms mixing  $S$ and right-handed neutrinos $\bar{\nu}$ in $\bar{\ell}$ appear:
\begin{equation}\label{MassYukawa}
    (m_D)_{ia} = \lambda^S_{ia}\,v_R.
\end{equation}
It is convenient to diagonalize this by the singular value decomposition,
\begin{equation}\label{SVD}
    m_D = U D_S V^\dagger,
\end{equation}
where $U$ and $V$ are unitary and $D_S$ is real diagonal with entries ${D_S}_i = M_{S_i}$. 
We perform transformations $S \to U^* S$ and $\bar{\nu}\to V^* \bar{\nu}$, so that the Dirac mass term becomes diagonal with eigenvalues $M_{S_i}$. The physical states are Dirac fermions $\Psi_{S_i} = \left(S_i, ~i \sigma^2\bar{\nu}^*_i \right)^t$ with masses $M_{S_i}$. In the following, we denote the physical heavy Dirac states $\Psi_{S_i}$ simply by $S_i$. With this convention, the decay channels relevant for the asymmetry are $S_i\to\bar e_a^\dagger W_R^\dagger$ and $S_i\to \ell_b^\dagger H_L^\dagger$ and their charge-conjugate decays, where $\bar{e}$ are right-handed charged leptons. 
We also rotate $\ell$ as $\ell\to V\,\ell$, so that the parity transformation $\ell \rightarrow \bar{\ell}^*$ does not involve $V$.
With this choice of basis, the $H_L$ Yukawa coupling is given by
\begin{equation}
\label{eq:Adef_massdiag}
   {\cal L} \supset  - S_i y_{ia}^S H_L \ell_a + {\rm h.c.},\qquad
   y_{ia}^S \equiv \frac{1}{v_R}\left(A D_S\right)_{ia},\qquad
   A \equiv U^\dagger U^*. 
\end{equation}
$A$ is symmetric and unitary. The coupling $y_{ia}^S$ is in general not diagonal or real.

\subsection{Charged-lepton flavor mixing}\label{subsec:flavor}
\begin{figure}[htb]
\centering
\includegraphics[width=0.9\linewidth]{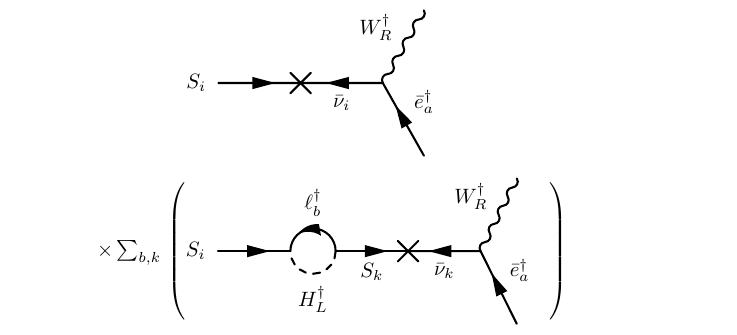}
\caption{The interference between the tree-level and one-loop diagrams contributing to 
$\epsilon_{ia}$. The loop sum runs over the intermediate singlet $S_k$ and the lepton flavor $b$. Arrows represent two-component fermion chirality flow.} 
\label{fig:Feynman}
\end{figure}

$S$ couples to right-handed charged leptons $\bar{e}$ and $W_R$ through its $\bar{\nu}$ component. CP violation comes from the quantum corrections by $y^S$ to the decay $S_i \to \bar{e}_a^\dagger W_R^\dagger$ shown in figure~\ref{fig:Feynman}, and the flavor structure of the charged-lepton sector affects the generation of lepton asymmetry.

We may neglect the SM lepton masses since they are much smaller than $M_S$. Then, if all of $\bar{e}_a$ are purely from $\bar{\ell}_a$, we may always choose the basis where only the diagonal gauge coupling between $S_i$ and $\bar e_a$ is nonzero, so $a=i$ in the upper panel of figure~\ref{fig:Feynman}. On the other hand, the CP violation could come from an intermediate singlet $S_k$ with $k\neq i$ in the lower panel of figure~\ref{fig:Feynman}, so there is no CP-violating interference between two diagrams, and the CP-violating decay is absent at one loop.
However, in the minimal Higgs model, $\bar{e}_a$ are generically a mixture of the charged component in $\bar{\ell}$ and another charged lepton in a different gauge representation, and CP violation may arise.

Let us consider three generations of $\ell$ and $\bar{\ell}$ as well as one generation of $E$ and $\bar{E}$ with Yukawa couplings and a mass,
\begin{equation}\label{eq:electronYukawa}
\mathcal{L} \supset -y^{\bar{e}}_{a} \ell_a H_L \bar{E} - {y_{a}^{\bar{e}\,*}}\bar{\ell}_a H_R E - M_{E}\bar{E} E +\text{h.c.},
\end{equation}
where $M_E$ is real due to parity symmetry.
After $SU(2)_R$ is broken,
\begin{equation}\label{electronUV}
\mathcal{L} \supset -y^{\bar{e}}_{a} \ell_a H_L \bar{E} - {y_{a}^{\bar{e}\,*} v_R }\bar{e}_a'E - M_{E}\bar{E} E +\text{h.c.},
\end{equation}
where $\bar{e}'$ are charged components of $\bar{\ell}$.

The relevant Dirac mass matrix can be expressed by the $1\times 4$ matrix 
\begin{equation}\label{eq:lepMassMatrix}
    \mathcal{M} = \left(v_R y^{\bar{e}\,*}_1,v_R y^{\bar{e}\,*}_2,v_R y^{\bar{e}\,*}_3,M_E \right).
\end{equation}
This can be diagonalized on the right by the $4\times 4$ unitary matrix $U^{\bar{e}}$ such that $\mathcal{M} U^{\bar{e}} = \left(0,0,0,M_H\right)$, where $M_H = \sqrt{v_R^2 \left\lVert y^{\bar{e}} \right \rVert^2 + M_E^2}$ with $\left\lVert y^{\bar e}\right\rVert \equiv \sqrt{\sum_{a=1}^{3} \left|y^{\bar e}_a\right|^2}$. $U^{\bar e}$ maps the mass eigenstates into the interaction eigenstates

\begin{equation}
    U^{\bar e} \begin{pmatrix}
        \bar{e}_1\\
        \bar{e}_2\\
        \bar{e}_3 \\
        \bar{e}_H
    \end{pmatrix}
    =
    \begin{pmatrix}
        \bar{e}'_1\\
        \bar{e}'_2\\
        \bar{e}'_3\\
        \bar{E}
    \end{pmatrix},
\end{equation}
where $\bar{e}_H$ has a mass of $M_H$ and $\bar{e}_1$, $\bar{e}_2$, and $\bar{e}_3$ are massless. We are interested in the upper-left $3\times3$ submatrix of $U^{\bar e}$ relating the interaction eigenstates $\bar e'_a$ to the light mass eigenstates $\bar e_a$. We denote this submatrix as $\tilde U^{\bar e}$. 
At scales below $M_H$, we integrate out the Dirac pair $(E,\bar e_H)$, giving an effective Lagrangian term
\begin{equation}\label{eq:EFT2Gen}
        \mathcal{L}_{\text{EFT}} = -\frac{v_R}{M_H} y^{\bar{e}}_a y^{\bar{e}\,*}_b \ell_a H_L \bar{e}_b +\mathrm{h.c.}
\end{equation} 
We identify the SM Yukawa coupling at scales below $M_H$ as $Y^{\mathrm SM}_{ab} = {v_R} y^{\bar{e}}_a y^{\bar{e}\,*}_b / {M_H}$. Since this effective coupling is rank-1, we identify the nonzero eigenvalue with the $\tau$ Yukawa coupling $y_{\tau}$, adopting the PDG central value~\cite{ParticleDataGroup:2024cfk}. In a complete three-generation charged-lepton sector, additional vector-like pairs are introduced to generate the electron and muon Yukawa couplings, but their contributions to the CP-violating decay of $S$ are negligible because of the small electron and muon Yukawa couplings.

If $M_E$ is taken to be small enough such that the right-handed charged-lepton mixing is $\lvert\tilde{U}^{\bar{e}}\rvert \sim \mathcal{O}(1)$, $\bar{e}_H$ can become lighter than $S_i$. In that limit, the inclusive CP-violating decay asymmetry cancels by unitarity. As discussed in Section \ref{subsec:CPviol}, the per-channel CP asymmetry is proportional to $\operatorname{Im}\left[U^{\bar e}_{ia}\,U^{\bar e\,*}_{ka} y^S_{ib} y^{S\,*}_{kb}\right]$, summed over intermediate singlet index $k\neq i$ and lepton index $b$. The physically relevant inclusive asymmetry sums over all kinematically accessible final charged leptons labeled by $a$. If $M_H\ll M_{S_i}$ then $\bar e_H$ is open and the sum runs over the full unitary matrix, implying that $\sum_{a}U^{\bar e}_{ia}U^{\bar e\,*}_{ka}=\delta_{ik}$, where $\delta$ is the Kronecker delta. For $k\neq i$ the prefactor vanishes, while for $k=i$ the remaining quantity is real, and so the inclusive CP-odd invariant vanishes. Thus, no net lepton asymmetry is generated in the limit where the entire right-handed charged-lepton sector is light and the mixing is fully unitary. We therefore require the heavy state to be kinematically inaccessible, $M_H \gtrsim M_{S_i}$.

This kinematic requirement does not parametrically suppress the inclusive $W_R^{(*)}$-mediated decay rate since a light charged-lepton combination remains accessible in the final state. However, when $M_{S_i}/v_R \gtrsim \lVert y^{\bar e} \rVert$, it suppresses the non-unitary flavor overlap $\sum_{a}U^{\bar e}_{ia}U^{\bar e\,*}_{ka}-\delta_{ik}$  that enters the CP-violating source. From Eq.~\eqref{eq:lepMassMatrix}, the projection of the physical charged-lepton eigenstates onto the interaction states parametrically follows $\theta_{\bar e}  \sim v_R \lVert y^{\bar e} \rVert/M_H$ for $\theta_{\bar e} \ll 1$. Taking the largest non-unitary flavor overlap consistent with $M_H \gtrsim M_{S_i}$ implies that $\theta^2_{\bar e}  \sim y_{\tau} v_R/M_{S_i}$. In our analysis with a right-handed neutrino mass scale $M$, we account for this by suppressing the source by the charged-lepton non-unitarity factor,
\begin{align}
    \theta^2_{\bar e}= {\rm min}\left(1, \frac{y_{\tau}v_R}{M} \right).
\end{align}

\section{CP violation}\label{sec:CPViolDyn}

\subsection{Right-handed neutrino decays}
The branching fraction for $S_i \to \bar{e}_a^\dagger W_R^\dagger$, and thus the physical CP violation associated with the decay of $S_i$, depends on the mass of $S_i$, $M_{S_i}$, and the mass of $W_R$, $M_{W_R} = g_{2,R}\,v_R/\sqrt 2$. By parity symmetry, the $SU(2)_R$ gauge coupling $g_{2,R}$ is equal to the $SU(2)_L$ gauge coupling at the $SU(2)_R$ breaking scale.

If $M_{S_i}$ is greater than the mass of $W_R$, then $S_i$ will decay into an on-shell $W_R$ and the off-shell $W_R$-mediated decays can be neglected.  The relevant partial decay widths when $M_{S_i} > M_{W_R}$ are
\begin{equation}\label{decay}
    \begin{aligned}
        \Gamma^{W_R}_{ia}
        &=
        \left|\tilde U^{\bar e}_{ia}\right|^2
        \frac{g_{2,R}^2 M_{S_i}}{64\pi}
        \left(2+\frac{M_{S_i}^2}{M_{W_R}^2}\right)
        \left(1-\frac{M_{W_R}^2}{M_{S_i}^2}\right)^2,
        \\
        \Gamma^{H_L}_{ib}
        &=
        \frac{\left|y^S_{ib}\right|^2}{16\pi}M_{S_i}.
    \end{aligned}
\end{equation}

If instead $M_{S_i} \leq M_{W_R}$, then $S_i$ cannot decay into an on-shell $W_R$. There is an off-shell $W_R$-mediated leptonic channel  $S_{i}\to\bar{e}_a^\dagger \bar{\nu}_k^\dagger \bar{e}_b$.
%We keep track of their generational indices as different combinations contribute distinctly to the lepton asymmetry.
There is also an off-shell $W_R$-mediated hadronic channel $S_{i}\to\bar{e}_a^\dagger \bar{u} \bar{d}^\dagger$, with $\bar{u}$ and $\bar{d}$ being massless right-handed SM up-type and down-type quarks respectively. At the scale $M_{W_R}$, integrating out $W_R$ gives a four-fermion operator with a coefficient proportional to $g_{2,R}^2/M_{W_R}^2$. The three-body widths therefore scale as $g_{2,R}^4 M_{S_i}^5/M_{W_R}^4$. The unpolarized decay widths at tree level for the leptonic and hadronic channels are
\begin{equation}\label{eq:decayOff}
    \begin{aligned}
    \Gamma^{\mathrm{Lep}}_{iakb}&=\frac{g_{2,R}^4\left|\tilde U^{\bar e}_{ia}\tilde U^{\bar e\,*}_{kb}\right|^{2}}{6144\,\pi^{3}}\,\frac{M_{S_i}^{5}}{M_{W_R}^{4}}\,F\!\left(\frac{M_{S_k}^{2}}{M_{S_i}^{2}}\right)\Theta\!\left(M_{S_i}-M_{S_k} \right),\\
    \Gamma^{\mathrm{Had}}_{ia}&=N_q\,\frac{g_{2,R}^4\left|\tilde U^{\bar e}_{ia}\right|^{2}}{2048\,\pi^{3}}\frac{M_{S_i}^{5}}{M_{W_R}^{4}},
    \end{aligned}
\end{equation}
where $F(x)= 1 - 8x + 8x^{3} - x^{4} - 12 x^{2}\ln x$ and $\Theta(x)$ is the Heaviside step function. We sum over three light-quark generations. Since the hadronic channel dominates the off-shell $W_R$-mediated decay, in numerical analysis we neglect the leptonic contribution.

\subsection{CP-violation parameter}\label{subsec:CPviol}
As discussed above,
$S$ decays via two channels,
one mediated by a massive $W_R$ boson and the other mediated by the massless $H_L$. We refer to the decay widths for each respective channel as $\Gamma(S_i \to \bar{e}_a^\dagger W_R^{(*)\dagger})$ and $\Gamma(S_i \to \ell_b^\dagger H_L^\dagger)$. We define the CP asymmetry per decay channel to be
\begin{equation}\label{eq:channelCP}
  \epsilon_{ia}\equiv \frac{\Gamma(S_i \to \bar{e}_a^\dagger W_R^{(*)\dagger}) - \Gamma\left(S^\dagger_i \to \bar{e}_a W_R^{(*)}\right)}{\Gamma(S_i \to \bar{e}_a^\dagger W_R^{(*)\dagger}) + \Gamma \left(S^\dagger_i \to \bar{e}_a W_R^{(*)}\right)}.
\end{equation}
This term represents the CP asymmetry normalized per decay channel and corresponds to the absorptive part of the interference diagram between the tree-level and one-loop processes, as seen in figure~\ref{fig:Feynman}. Here, the indices keep track of the singlet index of the initial $S_i$ and the lepton index of the final charged state.

We find (Appendix \ref{sec:AppendixCP}) that Eq.~\eqref{eq:channelCP}
can be expressed as
\begin{equation} \label{eq:eps_final}
    \begin{aligned}
    \epsilon_{ia} 
    &= \frac{1}{\left|\tilde U^{\bar e}_{ia}\right|^2}  \sum_{k\neq i}  \operatorname{Im} \left(\tilde U^{\bar e}_{ia} {\tilde U^{\bar e}_{ka}{}}^* \mathbf{Y}_{ik}\right) f_{ik},
    \end{aligned}
\end{equation}
to leading order in the Yukawa coupling and the mixing matrix $\tilde{U}^{\bar{e}}$. Here the charged-lepton index $b$ is summed inside as $\mathbf{Y}_{ik}=\sum_b y^{S}_{ib}\,y^{S\,*}_{kb}$, and the intermediate singlet index $k\neq i$ is summed over. We define
\begin{equation}\label{eq:loopFactor}
    f_{ik} \equiv \frac{1}{16 \pi}\;
    \frac{M_{S_i}M_{S_k}\left(M_{S_i}^2 - M_{S_k}^2 \right)}
    {\left(M_{S_i}^2 - M_{S_k}^2 \right)^2 + \left(M_{S_k}\Gamma^{\mathrm{Tot}}_k\right)^2},
\end{equation}
where $\Gamma^\mathrm{Tot}_k$ is the total decay width of the intermediate $S_k$, which we compute at tree level. We take the total decay width to be dominated by the tree-level $H_L$ decay channel. Importantly, this form does not diverge in the limit $M_{S_i}\to M_{S_k}$, since the intermediate propagator is regulated by the Breit--Wigner resummation. The resonant enhancement peaks when $|M_{S_i}^2-M_{S_k}^2|\sim M_{S_k}\Gamma_k^{\mathrm{Tot}}$~\cite{Dev:2017wwc,Garbrecht:2011aw,Garny:2011hg}. 

The physical CP violation relevant for leptogenesis is basis invariant and is controlled by the relative complex misalignment between the singlet Yukawa and the right-handed charged-lepton mixing. Field redefinitions can move phases between $\tilde U^{\bar e}$ and $\mathbf{Y}=y^S y^{S\dagger}$, but they cannot remove the invariant phase appearing in Eq.~\eqref{eq:eps_final}.

\subsection{Resonant subspace and flavor alignment}
The lepton asymmetry generated by the decay of the right-handed neutrinos depends on their mass differences. We work in the three-generation theory throughout but focus on the resonant subspace, in which $S_1$ and $S_2$ may be quasi-degenerate and $S_3$ is non-degenerate. We parameterize the degenerate pair by a central mass $M$ and splitting $\Delta M$,
\begin{equation}
    M_{S_1} = M-\frac{1}{2}\Delta M,~~ 
    M_{S_2} = M+\frac{1}{2}\Delta M,
\end{equation}
so that
\begin{equation}
    M=\frac{1}{2}(M_{S_1}+M_{S_2}), \qquad \Delta M = M_{S_2}-M_{S_1}.
\end{equation}
Let $\Delta M^2 \equiv M_{S_2}^2-M_{S_1}^2$, generally assuming the mass hierarchy of $M_{S_2}\gtrsim M_{S_1}$.

It is useful to first see what would happen in a strict two-generation theory. For the degenerate subspace, we can parameterize a $2\times 2$ symmetric unitary matrix $A$ with an angle $\theta$ and phase $\alpha$ up to an unphysical global phase. With the choice 
\begin{equation}\label{eq:Aparam}
    A =
    \begin{pmatrix}
        e^{i\alpha}\cos{\theta} & i\sin{\theta}  \\
        i\sin{\theta} &e^{-i\alpha}\cos{\theta}
    \end{pmatrix},
\end{equation}
it follows that 
\begin{equation}
    \mathbf{Y}=\frac{1}{v_R^2}
    \begin{pmatrix}
        M_{S_1}^2 \cos^2{\theta} +M_{S_2}^2 \sin^2{\theta} & -ie^{-i\alpha}\Delta M^2\sin{\theta} \cos{\theta} \\
        ie^{i\alpha}\Delta M^2\sin{\theta} \cos{\theta}& M_{S_1}^2 \sin^2{\theta} +M_{S_2}^2 \cos^2{\theta}
    \end{pmatrix}.
\end{equation}
Since $\mathbf{Y}_{12}$ and $\mathbf{Y}_{21}$ are proportional to $\Delta M^2$, the resonant enhancement of $\epsilon_{i a}$ is suppressed in the degenerate mass limit $\Delta M^2 \to 0$,
even though the factor in Eq.~\eqref{eq:loopFactor} is maximized. 

Returning to the three-generation theory,  $\mathbf{Y}$ is a $3\times 3$ matrix. Even with $S_1$ and $S_2$ being degenerate, if the third singlet is non-degenerate, $\mathbf{Y}$ is not generically proportional to the identity. The diagonal mass matrix of the three singlets is $D_S \simeq M\, \mathrm{diag}(1,1,\xi)$ with $\xi \neq 1$ parameterizing the non-degeneracy of $S_3$. Then, $\mathbf{Y}$ takes the form
\begin{equation}
\label{eq:Y_3gen}
    \mathbf{Y} \simeq \frac{M^2}{v_R^2}\,\mathbf{I}_3 + \frac{M^2}{v_R^2}\left(\xi^2 - 1 \right)  A\,\mathrm{diag}(0,0,1) \,  A^\dagger,
\end{equation}
with $\mathbf{I}_3$ being the identity in the three-generation $S$-flavor space.
Consequently, $\mathbf{Y}$ is generically not aligned with the heavy-singlet mass matrix in the three-generation case. Resonant enhancement is thus controlled by the small mass splitting of the nearly degenerate pair $S_1$ and $S_2$ while the non-degenerate spectator $S_3$ induces an off-diagonal element of $\mathbf{Y}$ in the degenerate subspace. 

We take $\xi=M_{S_3}/M>1$, so that $S_3$ is heavier than the resonant pair. This ensures that the washout of $B-L$ by the interaction of $S_3$ is negligible during relevant leptogenesis epoch by $S_{1,2}$. For thermal leptogenesis, this also ensures that the asymmetry created by $S_3$ is washed out by $S_{1,2}$. For non-thermal leptogenesis, we additionally assume that the inflaton directly decays into $S_{1,2}$, but not to $S_3$.
%We treat $S_3$ only as a flavor spectator and neglect its direct contribution to the generated asymmetry and washout.
%In the thermal analysis, we additionally assume that $S_3$ is sufficiently heavy that its population and inverse decays are suppressed during the relevant $S_1$--$S_2$ leptogenesis epoch.
%which require $S_3$ to be sufficiently heavy such that its inverse decays are suppressed during the relevant leptogenesis epoch. For thermal leptogenesis, this also ensures that asymmetry created by $S_3$ is washed out by $S_{1,2}$, although  
With this ordering of the masses, the second term in Eq.~\eqref{eq:Y_3gen} is positive semi-definite, and hence
\begin{equation}
    \label{eq:benchmark}
    \mathbf Y_{11},\,\mathbf Y_{22}\geq \frac{M^2}{v_R^2}.
\end{equation}

Let us derive $\epsilon_{ia}$ that maximizes baryon asymmetry, which is proportional to $\epsilon_{ia}$ and the branching ratio of the decay of $S_i$ into $\bar{e}_a+ W_R \propto |\tilde{U}_{ia}|^2/\mathbf{Y}_{ii}$.
We first focus on $i=1$, for which the possible resonance comes from $k=2$.
Since $\mathbf{Y}= y^S y^{S\dagger}$, the Cauchy--Schwarz inequality implies $\lvert \mathbf{Y}_{12}\rvert^2 \le \mathbf{Y}_{11}\mathbf{Y}_{22}$, so
\begin{align}
    \label{eq:eps_bound}
    |\epsilon_{1 a}| \leq\frac{|\tilde U^{\bar e}_{2a}|}{|\tilde U^{\bar e}_{1a}|}\sqrt{\mathbf{Y}_{11}\mathbf{Y}_{22}}\left|f_{12}\right| \simeq \frac{|\tilde U^{\bar e}_{2a}|}{|\tilde U^{\bar e}_{1a}|} \sqrt{\Gamma_{1}^{H_L} \Gamma_{2}^{H_L}}\frac{M\Delta M^2 }
    {\left(\Delta M^2 \right)^2 + \left(M \Gamma^{\mathrm{Tot}}_2\right)^2},
\end{align}
where $\Gamma_{i}^{H_L} = \sum_b \Gamma^{H_L}_{ib} = \mathbf{Y}_{ii}M_{S_i}/(16\pi)$.
 Although $|\epsilon_{1a}|$ is larger for larger $\Gamma_{1}^{H_L}$, large $\Gamma_{1}^{H_L}$ means that the branching ratio of the decay of $S_1$ into $\bar{e}_a + W_R$ becomes smaller, suppressing the baryon asymmetry produced by $S_1$ (larger $\Gamma_{1}^{H_L}$ also leads to stronger washout for thermal leptogenesis).
Therefore, the baryon asymmetry is parametrically maximized when $\Gamma^{H_L}_1\propto \mathbf Y_{11}$ is near its minimum allowed scale. For $\xi>1$, Eq.~\eqref{eq:Y_3gen} implies
$\mathbf Y_{11}\geq M^2/v_R^2$, so we take $\mathbf Y_{11}\sim M^2/v_R^2$, up to an order-one factor.

We choose the basis for the light right-handed charged leptons such that only $\bar e_1$ couples directly to $S_1$ and $W_R$, for which $\tilde U^{\bar e}_{11} = \mathcal O(1)$ and only $a=1$ contributes to asymmetry. $\tilde U^{\bar e}_{21}$ is then generated by the mixing between the singlet and the $SU(2)_R$ doublet charged leptons and is hence parametrically suppressed by $\theta^2_{\bar e}$.
Thus,
\begin{align}
\label{eq:epsilon_1a_max}
    |\epsilon_{11}^{\rm max}| \simeq \theta^2_{\bar e} \frac{M}{v_R} \sqrt{\frac{M\Gamma_{2}^{H_L}}{16\pi} }\frac{M\Delta M^2 }
    {\left(\Delta M^2 \right)^2 + \left(M \Gamma^{\mathrm{Tot}}_2\right)^2}
\end{align}
is the estimate for maximal $|\epsilon_{11}|$.
The optimal value of $\Gamma_{2}^{H_L}$ may depend on whether leptogenesis occurs thermally or non-thermally, which we discuss in the next section.

For $i=2$, only $\epsilon_{21}$ can be resonantly enhanced since we take the basis where only $\bar{e}_1$ couples to $S_1$. $\epsilon_{21}$ that maximizes baryon asymmetry is also given by Eq.~\eqref{eq:epsilon_1a_max}, with replacing $\Gamma_2^{H_L,{\rm Tot}}$ with $\Gamma_1^{H_L,{\rm Tot}}$. The sign of $\epsilon_{21}$ is the same as that of $\epsilon_{11}$, so there is no cancellation between the asymmetries produced by $S_1$ and $S_2$.

In the next section, we take small $\Delta M/M$ to enhance baryon asymmetry. Vetoing fine-tuning, $\Delta M$ cannot be arbitrarily small because of radiative corrections by Yukawa interactions. 
The one-loop wavefunction renormalization of $S$ is proportional to ${\mathbf Y}$.  After a unitary rotation and rescaling to canonicalize the kinetic term of $S$, a contribution to $\Delta M /M$ proportional to $\sqrt{\left(\mathbf Y_{11} - \mathbf Y_{22}\right)^2 + 4|\mathbf Y_{12}|^2}$ is generated, which gives
\begin{equation}
    \label{eq:SingletWavefunctionCorrection}
    \left.\frac{\Delta M}{M}\right|_S \sim \frac{1}{16 \pi^2}\left(\frac{M}{v_R}\right)^2\,\ln{\left(\frac{\Lambda}{v_R}\right)},
\end{equation} where $\Lambda$ is the UV scale.
The charged-lepton sector provides an additional independent source of mass splitting. Following Ref.~\cite{Casas:1999tp}, wavefunction renormalization of $\bar{\ell}$ from the Yukawa coupling in Eq.~\eqref{eq:electronYukawa} induces a fractional splitting on the degeneracy of
\begin{equation}\label{eq:WavefunctionCorrection}
    \left.\frac{\Delta M}{M}\right|_\ell
    \simeq
    \frac{3(y^{\bar{e}})^2}{32\pi^2} \,\ln{\left(\frac{\Lambda}{v_R}\right)} \geq \frac{3y_{\tau}^2}{32\pi^2} \,\ln{\left(\frac{\Lambda}{v_R}\right)}.
\end{equation}
For numerical estimates, we take $\Lambda \sim M_{\mathrm{Pl}}$. The lowest symmetry-breaking scale we consider to be viable is $v_R\sim 10^{4}\,\mathrm{GeV}$, for which the charged-lepton contribution gives $\Delta M/M \gtrsim 3\times 10^{-5}$. The singlet wavefunction renormalization becomes dominant for $M/v_R \gtrsim y_{\tau}$, above which the radiative floor grows in proportion to $(M/v_R)^2$.

We have focused on the case with three singlets $S_{1,2,3}$. The observed neutrino masses can be explained by two singlets $S_{1,2}$ without $S_3$. This case, however, leads to ${\mathbf Y} \propto \mathbf{I}_2 $ in the limit $\Delta M/M \ll 1$, so the resonant enhancement is absent.
In Section~\ref{sec:Asymm}, we also analyze  $\Delta M/M \sim 1$, which is applicable to the two-singlet case.

\section{Baryon asymmetry}\label{sec:Asymm}

A net baryon yield can be generated by converting a net lepton asymmetry into baryon number through $B\!-\!L$-conserving electroweak sphaleron processes. For a species $X$ with number density $n_X$, we work with yields $Y_X\equiv n_X/s$, where $s$ is the entropy density of the Universe. We use Maxwell--Boltzmann statistics for all number densities, linearization in chemical potentials for light species, and leading-order terms in the CP-violation parameters. Sphaleron processes fall out of equilibrium shortly after electroweak symmetry crossover, so we use~\cite{PhysRevD.42.3344}
\begin{equation}\label{eq:HarveyTurnerYield}
  Y_B = \frac{28}{79}\,Y_{\Delta(B-L)}.
\end{equation}
Here $Y_{\Delta(B-L)}$ denotes the $B\!-\!L$ charge stored in the SM plasma after the heavy singlet decays. The factor $28/79$ is the standard electroweak-sphaleron conversion from this conserved $B\!-\!L$ yield into baryon number in the SM plasma.
We take the observed baryon yield to be $Y_B^{\rm obs}\approx 8.66\times10^{-11}$~\cite{Planck:2018vyg}.

\subsection{Thermal leptogenesis}\label{subsec:therm}

In our setup $Y_{\Delta(B-L)}$ is sourced by the out-of-equilibrium decays of $S$, while SM interactions conserve $B\!-\!L$. In the Boltzmann system of Appendix~\ref{sec:AppendixBoltz}, we track the individual number-asymmetry yields $Y_{\Delta X}\equiv (n_X-n_{\bar X})/s$ for the light species $X$ and combine them into the final conserved $B\!-\!L$ yield.

The evolution of the singlet abundances and light-species asymmetries is governed by a system of Boltzmann equations. Because of the efficient $SU(2)_R\times U(1)_X$ interactions at high temperatures, the initial conditions for the singlets are given by the thermal ones, $Y_{S_i}(0)=Y_{S_i}^{\rm eq}(0)$ and $Y_{S_i^\dagger}(0)=Y_{S_i}^{\rm eq}(0)$, with zero initial asymmetries. We define 
$z\equiv M_{S_1}/T$ and $x_i\equiv M_{S_i}/M_{S_1}$. For any decay width $\Gamma_{ia}$ we use the Maxwell--Boltzmann thermal average
\begin{equation}
\langle \Gamma_{ia}\rangle=\Gamma_{ia}\,\frac{\mathcal K_1(x_i z)}{\mathcal K_2(x_i z)},
\end{equation}
with $\mathcal K_i$ being the modified Bessel functions of the second kind. We keep decays and inverse decays of $S$ and ignore two-to-two scatterings, which are subdominant for $T<M_{S_i}$. 
Following the standard first-order treatment~ in the CP-violation parameter~\cite{Buchmuller:2004nz}, we express the final baryon yield as a linear functional of the CP-violation parameters,
\begin{equation}\label{eq:kappaDef_sec4}
Y_B(\infty) = \sum_{i,a}\epsilon_{ia}\,\kappa_{ia}\left[Y_{S_i}^{\mathrm{eq}}(0) + Y_{S_i^{\dagger}}^{\mathrm{eq}}(0)\right],
\end{equation}
where $\kappa_{ia}$ is the efficiency matrix determined by the Boltzmann system in Appendix \ref{sec:AppendixBoltz}.

To obtain the bound on $Y_B$, we construct analytic upper bounds on the efficiency factor in the on-shell, $M_{S_i}>M_{W_R}$, and off-shell, $M_{S_i}\leq M_{W_R}$, regimes, across weak and strong washout~\cite{Buchmuller:2004nz}. We denote the total $W_R^{(*)}$-mediated source width into flavor $a$ by $\Gamma^{W_R(*)}_{ia}$. In the on-shell regime, $\Gamma^{W_R(*)}_{ia}=\Gamma^{W_R}_{ia}$, while in the off-shell regime 
%$\Gamma^{W_R(*)}_{ia}=\sum_{k,b}\Gamma^{\rm Lep}_{iakb}+\Gamma^{\rm Had}_{ia}$.
$\Gamma^{W_R(*)}_{ia}=\Gamma^{\rm Had}_{ia}$.
We define the corresponding branching fraction as $\operatorname{Br}^{W_R(*)}_{ia}\equiv \Gamma^{W_R(*)}_{ia}/\Gamma^{\rm Tot}_{i}$.

In the weak-washout regime, we neglect washout terms proportional to the asymmetries and set $Y_{S_i^-}=0$ at leading order, since it is itself sourced by the asymmetries. Using the Boltzmann system in Appendix~\ref{sec:AppendixBoltz} and the standard integral identity following from the CP-even singlet equation gives
\begin{equation}\label{eq:DI_kweak}
\kappa^{\rm weak}_{ia}\simeq\frac{56}{79}\operatorname{Br}^{W_R(*)}_{ia}.
\end{equation}
The same charge factor applies in both kinematic regimes. The $\bar e^\dagger W_R^{(*)\dagger}$ final state carries $B-L=-1$, while the  $\ell^\dagger H_L^\dagger$ final state carries $B-L=+1$. In particular, the quark pair in the hadronic three-body decay carries zero net $B-L$.

In the strong-washout regime, washout is dominated by the $H_L\ell$ interactions. We therefore parameterize the corresponding interaction strength by
\begin{equation}\label{eq:DI_K_def}
K_i\equiv\frac{\sum_a\Gamma^{H_L}_{ia}}{H(M_{S_i})}.
\end{equation}
Here, $H(T)\simeq1.66\sqrt{g_*}\,T^2/M_{\rm Pl}$ is the Hubble parameter during radiation domination. The $W_R^{(*)}$-mediated rates are retained in the CP-odd source through $\operatorname{Br}^{W_R(*)}_{ia}$, while their inverse decays are neglected in the washout terms.

We choose $\mathbf Y_{22}$ near its minimum allowed value in order to minimize washout. From Eq.~\eqref{eq:benchmark}, $\mathbf Y_{22}\geq M^2/v_R^2$, and we therefore take $\mathbf Y_{22}\sim M^2/v_R^2$, up to an order-one factor. This choice is also favored by the resonant CP asymmetry since at $\Delta M^2=M\Gamma^{H_L}_2$, one has $|\epsilon_{11}^{\rm max}|\propto\sqrt{\mathbf Y_{11}/\mathbf Y_{22}}$, so increasing $\mathbf Y_{22}$ does not enhance the resonant peak while it increases the washout rate. Together with $\mathbf Y_{11}\sim M^2/v_R^2$, this gives $K_1\sim K_2\sim K$, with
\begin{equation}\label{eq:DI_K_scaling}
K\simeq1\times\left(\frac{M}{10^8\,\mathrm{GeV}}\right)\left(\frac{10^{12}\,\mathrm{GeV}}{v_R}\right)^2.
\end{equation}

The CP-odd source is concentrated near $z=z_B$. We use the standard fit~\cite{Buchmuller:2004nz}
\begin{equation}\label{eq:DI_zB_fit}
z_B(K)=1+\frac{1}{2}\ln\left[1+\frac{\pi K^2}{1024}\left(\ln\frac{3125\pi K^2}{1024}\right)^5\right].
\end{equation}
For the quasi-degenerate resonant pair, the departure from equilibrium of each $S_i$ is controlled by its own interaction rate. Hence we evaluate the suppression for each contribution using $K_i \simeq K$. Away from the quasi-degenerate resonant regime, the analytic bound is dominated by the lighter state $S_1$, and we thus use $K_1$.
In the strong-washout limit, the efficiency matrix follows
\begin{equation}\label{eq:DI_kstrong_ratio}
\kappa^{\rm strong}_{ia}\simeq\kappa^{\rm weak}_{ia}\frac{Y^{\rm eq}_{S_i}(z_B)}{Y^{\rm eq}_{S_i}(0)}\simeq\kappa^{\rm weak}_{ia}\frac{\pi^2}{12\zeta(3)Kz_B(K)},
\end{equation}
where the appropriate washout parameter specified above is understood. We interpolate between the weak- and strong-washout limits using
\begin{equation}\label{eq:DI_kappa_tilde}
\widetilde{\kappa}_{ia}(K)\equiv\frac{\pi^2\kappa^{\rm weak}_{ia}}{12\zeta(3)Kz_B(K)}\left[1-\exp\left(-\frac{12\zeta(3)}{\pi^2}Kz_B(K)\right)\right].
\end{equation}
%In the off-shell region relevant for the analytic upper bound, the hadronic three-body width typically dominates the $W_R^*$-mediated source. We therefore use $\Gamma^{W_R(*)}_{ia}\simeq\Gamma^{\rm Had}_{ia}$ when evaluating the analytic approximation.

In the degenerate limit and $\mathbf{Y}_{11} \simeq \mathbf{Y}_{22}$, $S_1$ and $S_2$ give the same contribution to the baryon asymmetry,  so the maximal baryon asymmetry is given by
\begin{equation}\label{eq:YB_UB_def}
    \left|Y_B^{T,{\rm resonance}}\right|
    =
    2\,
    |\epsilon_{11}^{\rm max,T}|\,
    \widetilde{\kappa}\,
    Y_{S}^{\mathrm{eq}}(0).
\end{equation}
$\Delta M/M$ is chosen to be the larger of two quantities: the value that maximizes $\epsilon_{1a}$,
\begin{equation}
\label{eq:resonantCondition}
\Delta M^2 = M \Gamma_2^{H_L},
\end{equation}
and the radiative lower bound from Eqs.~\eqref{eq:SingletWavefunctionCorrection} and~\eqref{eq:WavefunctionCorrection}. Namely, we take the radiative floor to be the sum of the singlet- and charged-lepton-wavefunction contributions, and use the resonant splitting when it lies above this floor. Otherwise, we use the smallest technically natural mass splitting allowed.

Since $\Delta M/M\ll 1$ would require the introduction of flavor symmetry, we also consider a simpler case where $\Delta M/M = \mathcal O(1)$. Away from resonance, taking $\mathbf Y_{22}> \mathbf Y_{11}$ can enhance the CP asymmetry through $|\mathbf Y_{12}|^2\leq \mathbf Y_{11} \mathbf Y_{22}$. However, it also increases the $S_2$ Yukawa width and the associated washout. We have checked that allowing $\mathbf Y_{22}> \mathbf Y_{11}$ changes the resulting parameter-space boundary only by an $\mathcal O(1)$ factor. We therefore use the simple benchmark $\mathbf Y_{22}\simeq \mathbf Y_{11}\sim M^2/v_R^2$. Taking $\Delta M^2= M^2$, we obtain
\begin{equation}
    \label{eq:YB_nonresonance}
    \left|Y_B^{T,{\rm non-resonance}}\right|
    = \theta^2_{\bar{e}}
     \frac{M^2}{16\pi v_R^2 } \,
    \widetilde{\kappa}\,
    Y_{S}^{\mathrm{eq}}(0).
\end{equation}
Here the factor $(M/v_R)^2/(16\pi)$ gives the expected perturbative size of the self-energy CP asymmetry without resonant enhancement.

\begin{figure}[t]
    \centering
    \includegraphics[width=0.8\linewidth]{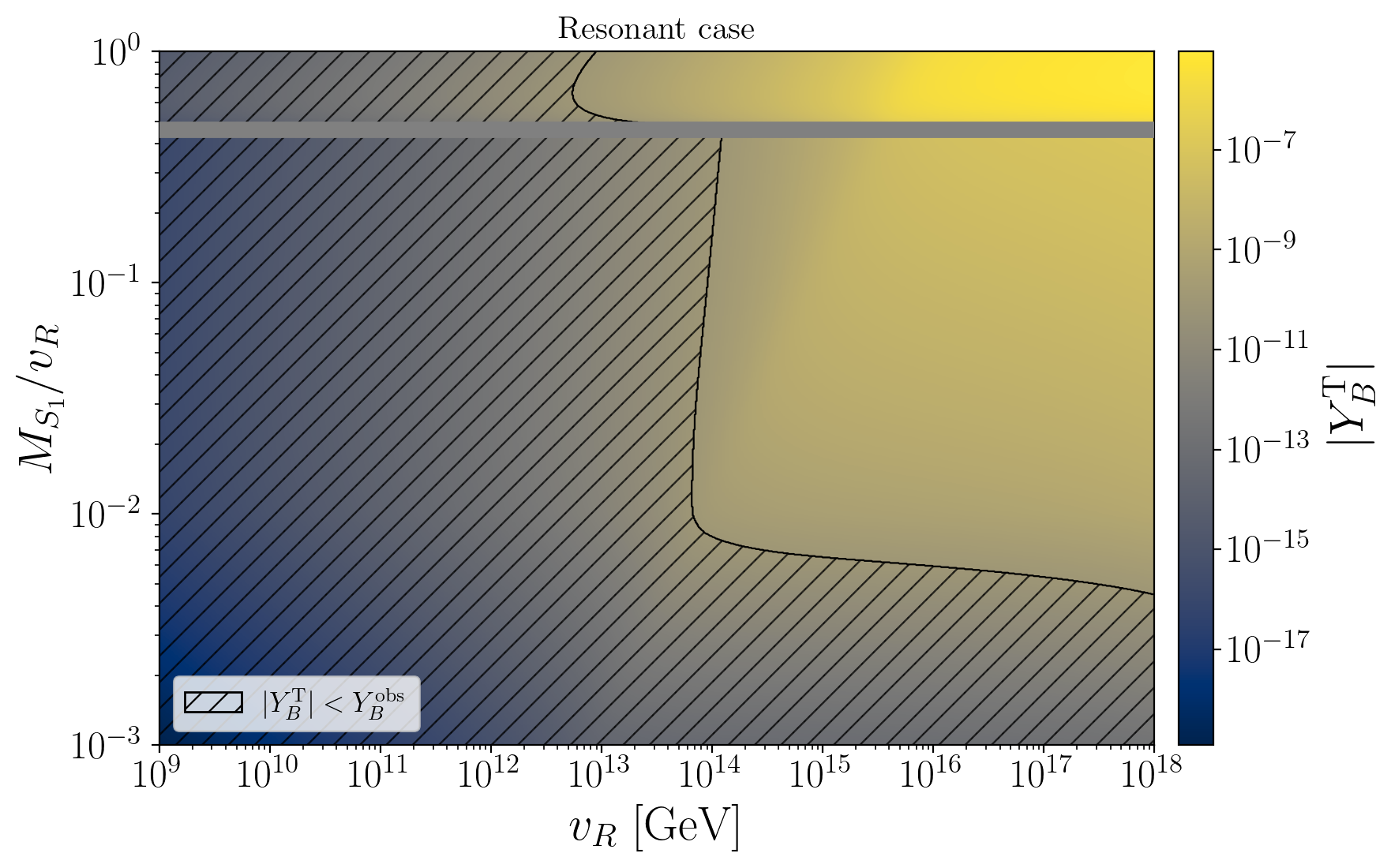}
    \includegraphics[width=0.8\linewidth]{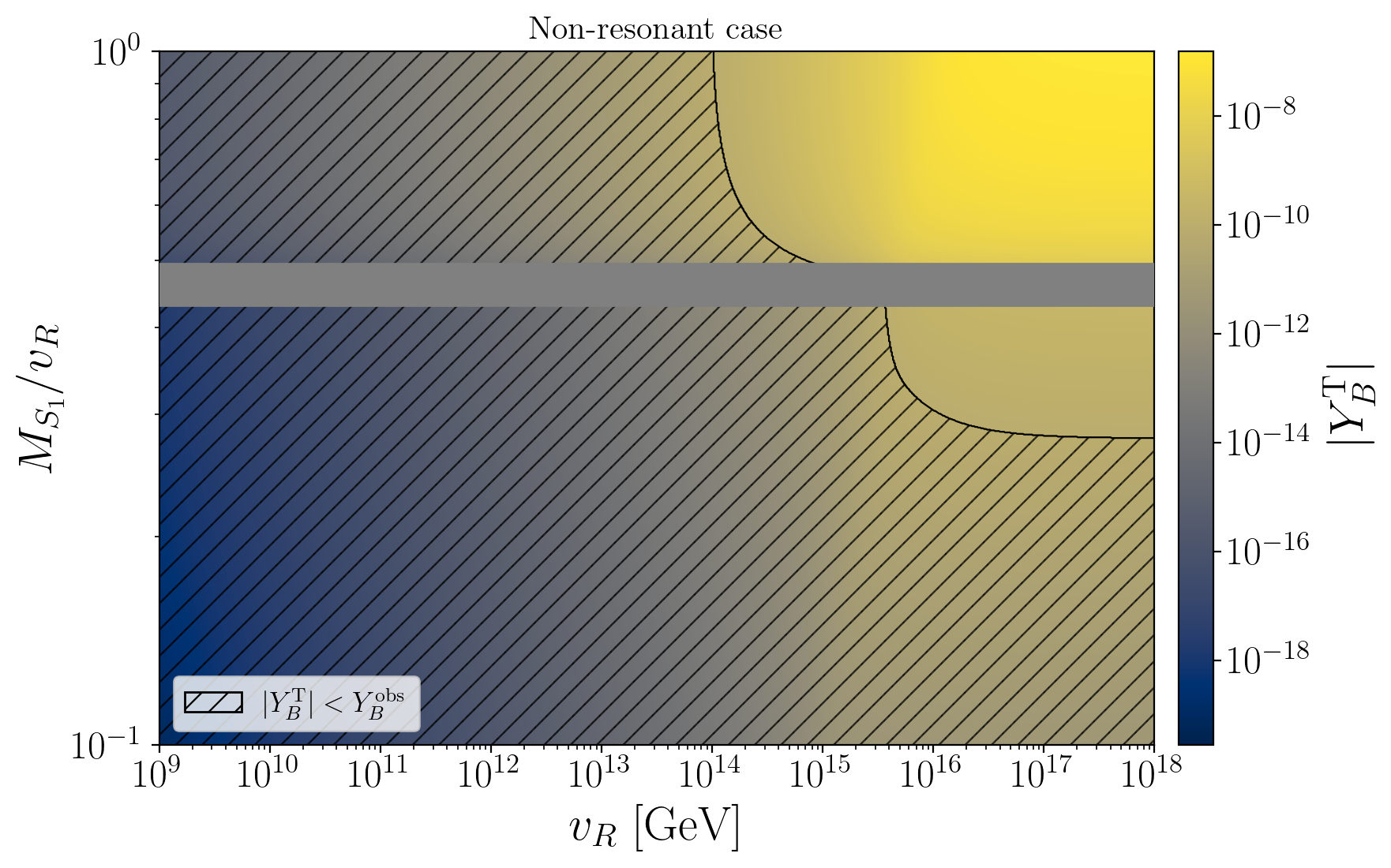}
    \caption{
    Constraints on thermal leptogenesis. Top: the resonant case with the minimal mass splitting allowed by radiative corrections. The contributions of the quasi-degenerate pair are included as in Eq.~\eqref{eq:YB_UB_def}, with the resonant self-energy factor evaluated using Eq.~\eqref{eq:loopFactor}. The fractional splitting $\Delta M/M$ is chosen to be the larger of the value required by the resonance condition and the radiative floor.
    Bottom: the non-resonant benchmark of Eq.~\eqref{eq:YB_nonresonance}. In both panels, the color scale shows the analytic estimate for the maximal $\left|Y_B^T\right|$. Hatched regions satisfy $\left|Y_B^T\right|<Y_B^{\rm obs}$ and therefore do not generate the observed baryon asymmetry. The gray horizontal band indicates the crossover between the on-shell and off-shell $W_R$ regimes.}
    \label{fig:YB_exclusion}
\end{figure}

Parameter points that cannot generate the observed baryon asymmetry ($\left|Y_B^{T}\right|<Y_B^{\rm obs}$) are shown in figure~\ref{fig:YB_exclusion}. Even with resonance, successful thermal leptogenesis requires $v_R\gtrsim 6\times 10^{12}\,\mathrm{GeV}$, while the non-resonant benchmark is more strongly constrained because the CP asymmetry is smaller.

Although it is difficult to directly probe the model by collider searches for new particles, this high $v_R$ is consistent with the scale suggested by the vanishing of the SM Higgs quartic coupling, $(10^9\text{--}10^{13})\,\mathrm{GeV}$~\cite{Hall:2018let}. By more precisely measuring the top quark mass and the strong coupling constant, we may narrow down the range of $v_R$ to test the consistency with our setup.

\FloatBarrier

\subsection{Non-thermal leptogenesis}\label{subsec:nontherm}
We turn to the scenario where the initial yield of $S_1$ is generated via the decay of an inflaton $\varphi$ (or any field that may dominate the universe) with mass $m_\varphi > 2M_{S_1}$, allowing for a non-thermally generated baryon asymmetry~\cite{Fukugita:1986hr,Asaka:1999yd,Bassett:2005xm}.
%We assume that reheating completes after $SU(2)_R$ breaking, requiring $T_{\mathrm{RH}}\lesssim v_R$. 
Our discussion does not depend on the detailed microphysics of the inflaton sector and the interaction leading to the  decay of the inflaton into right-handed neutrinos, but an example includes the decay by a coupling $\varphi S^2$. (We comment on a possible model dependence of the viable parameter space at the end of this section.)
In this work, ``non-thermal'' means that the singlet abundance is set by inflaton decay rather than by thermal production from the plasma, which provides a deviation from thermal equilibrium.  We focus on the case where $ T_{\rm test}\equiv \max(T_{\rm RH},T_{\rm EW})<M_{S_1}$, so that the singlet decay can be treated without taking into account thermal effects.%
\footnote{
Non-thermal leptogenesis is possible even for $T_{\rm test} >M_{S_1}$. Although $S_1$ is thermalized at $T=T_{\rm test}$ by the Yukawa interaction, since the initial abundance is set by the inflaton decay, $B-L$ can be produced while the abundance of $S$ decreases from the initial value to the thermal value. As long as $W_R$ exchange is ineffective at $T=T_{\rm test}$, $B-L$ is not washed out. 
}

The inflaton dominates the energy density before reheating, so at reheating temperature $T_{\mathrm{RH}}$, the energy density of the universe  is $\rho=\frac{\pi^2}{30}g_*T_{\mathrm{RH}}^4$. Taking the inflaton number density as $n_\varphi \simeq \rho/m_\varphi$, the inflaton yield at reheating is
\begin{equation}
Y_\varphi \equiv \frac{n_\varphi}{s}\simeq \frac{3}{4}\frac{T_{\mathrm{RH}}}{m_\varphi}.
\end{equation}
The non-thermally generated singlet yield from inflaton decay is then $Y^{\mathrm{NT}}_{S_i}=\mathrm{Br}_{\mathrm{infl}}Y_\varphi$, where $\mathrm{Br}_{\mathrm{infl}}$ is the average number of $S$ produced per inflaton decay. For example, when the inflaton dominantly decays into two $S$, $\mathrm{Br}_{\mathrm{infl}}=2$.

If reheating completes before the electroweak sphaleron process freezes out at $T= T_{\mathrm {EW}}\sim 100\,$GeV, then the standard source estimate from Eq.~\eqref{eq:HarveyTurnerYield} applies directly. However, if $T_{\mathrm{RH}} < T_{\mathrm{EW}}$, only the fraction of inflaton decays that occur before the plasma cools through the sphaleron-active regime can contribute to the final baryon asymmetry. The fraction inflaton decays that occur while the electroweak sphaleron process is still active is
\begin{equation}
    f_\varphi(T_{\mathrm{EW}})=
    \begin{cases}
        1, & T_{\mathrm{RH}} \ge T_{\mathrm{EW}},\\[6pt]
        \left(\dfrac{T_{\mathrm{RH}}}{T_{\mathrm{EW}}}\right)^4, & T_{\mathrm{RH}} < T_{\mathrm{EW}},
    \end{cases}
    \label{eq:fphi_piecewise}
\end{equation}
which encodes the scaling $H\propto T^4$ before the completion of reheating~\cite{Kolb:2003ke}.%
\footnote{
If the decay products of $S$ are not immediately thermalized, the radiation of the universe does not follow thermal distributions. For the low reheating temperatures we consider, thermalization occurs well before reheating completes~\cite{Harigaya:2013vwa}.
}

We parameterize the baryon asymmetry generated non-thermally as
\begin{equation}\label{eq:YB_UB_NT}
\left|Y_B^{\mathrm{NT}}\right|
\simeq
\frac{42}{79}\,| \epsilon_{11}^{\mathrm{max}}|\operatorname{Br}_{W_R^{(*)}}\,
\frac{T_{\mathrm{RH}}}{m_\varphi}\,\mathrm{Br}_{\mathrm{infl}}\,f_\varphi (T_\mathrm{EW}).
\end{equation}
In evaluating $|\epsilon^{\max}_{11}|$, we take the benchmark $\mathbf Y_{22}\simeq \mathbf Y_{11}\sim M^2/v_R^2$ and saturate $|\mathbf Y_{12}|^2\leq \mathbf Y_{11} \mathbf Y_{22}$ up to order-one factors. For a fixed mass splitting such that $\Delta M^2 > M^4/(16 \pi v_R^2)$, increasing $\mathbf Y_{22}$ can move $\Gamma^{H_L}_2$ closer to the value that maximizes the self-energy enhancement. We have checked that allowing this enhancement changes the boundary of the viable parameter space only by an $\mathcal O(1)$ factor, even before including the additional washout constraint associated with larger $\mathbf Y_{22}$. We therefore retain $\mathbf Y_{22}\simeq \mathbf Y_{11}$ as our benchmark.

The reheating temperature required to produce the observed baryon asymmetry is
\begin{equation}
    T_{\mathrm {RH}}=
    \begin{cases}
        \dfrac{79}{42}\dfrac{Y_B^{\rm obs}m_\varphi}{|\epsilon_{11}^{\mathrm{max}}|\operatorname{Br}_{W_R^{(*)}}\mathrm{Br}_{\mathrm{infl}}}, & T_{\mathrm{RH}} \ge T_{\mathrm{EW}},\\[12pt]
        \left(\dfrac{79}{42}\dfrac{Y_B^{\rm obs}m_\varphi}{|\epsilon_{11}^{\mathrm{max}}|\operatorname{Br}_{W_R^{(*)}}\mathrm{Br}_{\mathrm{infl}}}\,T_{\mathrm{EW}}^4\right)^{1/5}, & T_{\mathrm{RH}} < T_{\mathrm{EW}}.
    \end{cases}
    \label{eq:TRH_piecewise}
\end{equation}
For every point in the parameter region of interest, we first solve for the reheating temperature required to reproduce the observed baryon asymmetry, and then test whether that reheating temperature is consistent with our assumptions and the absence of washout.

Washout of $B-L$ is effective when the chain $\ell H_L \leftrightarrow S \leftrightarrow\bar e W_R^{-(*)}$ is fast enough to impose chemical equilibrium between the left-handed and right-handed lepton sectors. Thus, if either the $H_L$-mediated inverse decay or the $W_R^{-(*)}$-mediated inverse decay is inefficient, this washout channel is ineffective. By detailed balance, we exclude points in parameter space using the condition for effective washout
\begin{equation}
\frac{n^{\mathrm{eq}}_{S_i}}{n^{\mathrm{eq}}_{\ell_a}}
\left\langle \Gamma^{H_L}_{ia}\right\rangle > H(T_{\mathrm{test}}) \quad \mathrm{and}\quad \frac{n^{\mathrm{eq}}_{S_i}}{n^{\mathrm{eq}}_{\bar e_b}}
\left\langle \Gamma^{W_R(*)}_{ib}\right\rangle > H(T_{\mathrm{test}}),
\label{eq:rateID}
\end{equation}
where $n^{\mathrm{eq}}_X$ is the equilibrium number density of the species $X$. The thermal average is evaluated at $T_{\mathrm{test}}$. For $M_{S_i}<M_{W_R}$, we treat the off-shell $W_R$-mediated decay using the leading contact-interaction approximation, neglecting finite-width and threshold effects of the virtual $W_R$.
For $T_{\rm test}<M_{S_i}$, which we assume in our analysis, the washout rate by scattering is suppressed in comparison with that by decay.

Washout also occurs in processes without $S_1$ in the external states.
After integrating out both the singlet and the $W_R$, there is a scattering process, $\ell_a H_L \leftrightarrow \bar e_b\,\bar u^\dagger\,\bar d$ via a dimension-seven interaction
\begin{equation}
    \mathcal L_7
    \supset
    C_{aib}\,
    H_L
    \left(\ell_a^\dagger \bar\sigma^\mu \bar e_b\right)
    \left(\bar u^\dagger\bar\sigma_\mu\bar d\right)
    +\mathrm{h.c.},
\end{equation}
where
\begin{equation}
    C_{aib}
    =
    \frac{g_{2,R}^2}{4\sqrt{2}}\,
    \frac{\tilde U^{\bar e}_{ib}}{M_{W_R}^2}\,
    \frac{y^{S*}_{ia}}{M_{S_i}}.
\end{equation}
We sum over three quark colors and take the incoming $\ell_a$ and $H_L$ to follow the Maxwell--Boltzmann distribution. This gives the thermally averaged washout rate
\begin{equation}
    \left\langle
    \Gamma_{ab}^{\rm scatt}
    \right\rangle
    \simeq
    \frac{9}{128\pi^5}
    \left|\mathcal F_{ab}\right|^2
    \frac{T_{\rm test}^7}{v_R^6},
\label{eq:double_scatter}
\end{equation}
where
\begin{equation}
    \mathcal F_{ab}
    \equiv
    \sum_i
    \tilde U^{\bar e}_{ib}
    \frac{y^{S*}_{ia}v_R}{M_{S_i}}.
\end{equation}
The factor $\left|\mathcal F_{ab}\right|^2$ contains the sum
over the virtual singlet states and the dependence on the
charged-lepton flavor texture for a fixed external channel. The total
washout rate is obtained by summing over the kinematically accessible
external lepton flavors. We therefore define $\left|\mathcal F\right|^2 \equiv \sum_a\sum_{b\in{\rm light}} \left|\mathcal F_{ab}\right|^2$. In the phenomenological analysis, we take $\left|\mathcal F\right|^2=1$ as an unsuppressed benchmark. This choice is motivated by $y^S_{ia}v_R/M_{S_i} \sim \mathcal O(1)$ together with an order-one $\tilde U^{\bar e}_{ib}$, but smaller values of $\left|\mathcal F\right|^2$ can in principle arise from destructive interference among the virtual singlet contributions.
When the total scattering rate of the dimension-seven effective interaction exceeds the Hubble rate in its domain of validity, it acts as a washout process and exponentially damps any pre-existing $B-L$ asymmetry. We therefore include the region satisfying $\Gamma_{\rm scatt}\gtrsim H(T_{\rm test})$ as a hard washout exclusion.
In the parameter regions that are otherwise allowed, $T_{\rm test} \ll M_{W_R},M_{S_{i}}$, so the estimation of the scattering rate via the dimension-seven operator is valid.
Since ${\Gamma_{\rm scatt}}/{H} \propto T_{\rm test}^5/{v_R^6}$, this washout is mostly relevant at low $v_R$.

Our estimation of the baryon asymmetry assumes $\Gamma^{\mathrm{Tot}}_S > H(T_{\mathrm{EW}})$, so that the singlet decays occur while the electroweak sphaleron process is still active. We find that this condition is always satisfied in the viable parameter region.

\begin{figure}[th]
    \centering
    \includegraphics[width=0.72\linewidth]{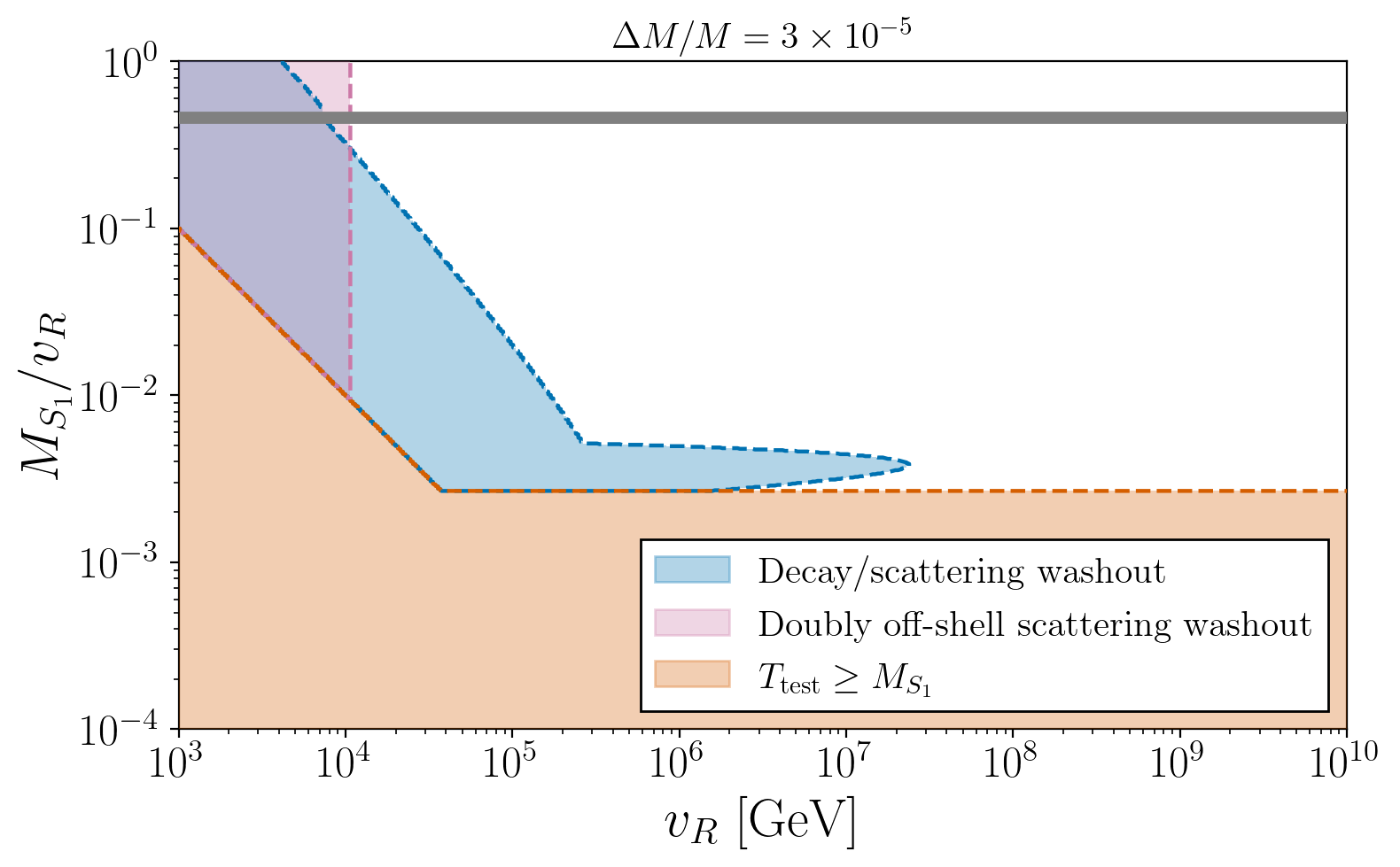}
    \includegraphics[width=0.72\linewidth]{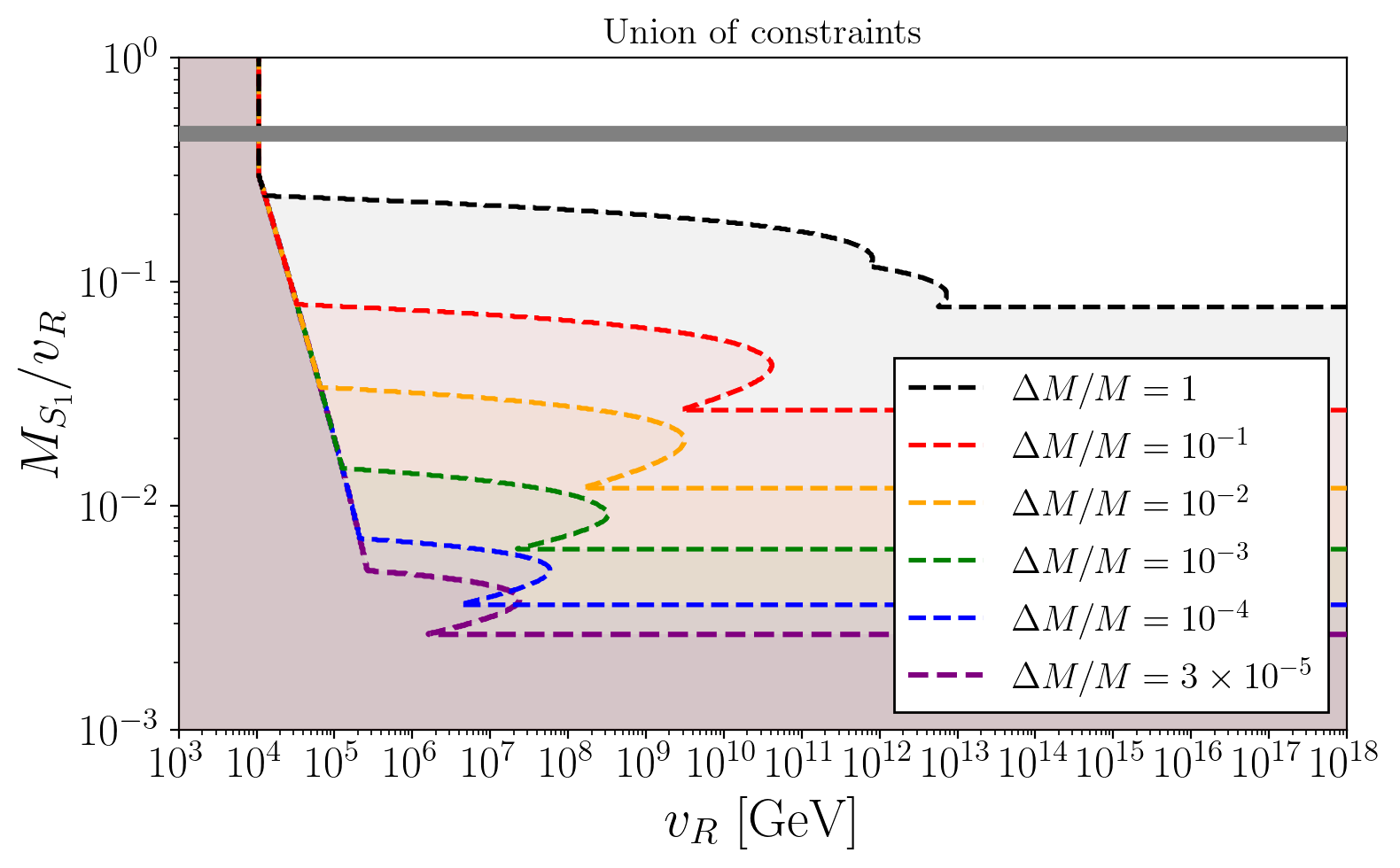}
    \caption{Constraints on $(v_R,M_{S_1}/v_R)$ for non-thermal leptogenesis. The gray horizontal band indicates the crossover between the on-shell and off-shell $W_R$ regimes. Top: decomposition of each constraint for $\Delta M/M=3\times10^{-5}$. 
    In the blue- and pink-shaded regions,
    inverse decays and doubly off-shell scattering wash out $B-L$, respectively. The orange-shaded region is inconsistent with the assumption that $T_{\rm test}<M_{S_1}$. Bottom: union of constraints for several fixed mass splittings.}
    \label{fig:nontherm}
\end{figure} 

The allowed regions are shown in figure~\ref{fig:nontherm}. We take $m_\varphi=2 M$ and $\mathrm{Br}_{\mathrm{infl}}=2$ in our numerical analysis. For the quasi-degenerate benchmarks, this is a near-threshold choice that maximizes the non-thermal singlet yield for fixed reheating temperature; for the non-resonant reference $\Delta M/M=1$, we take $m_\varphi=2M_{S_1}$. In these benchmarks, the decay $\varphi\to S_3S_3$ is kinematically forbidden. A larger inflaton mass reduces the yield and therefore requires a proportionally larger $T_{\rm RH}$.
Small $M_{S_1}$ requires larger $T_{\rm test}$ and is inconsistent with the assumption $T_{\rm test} < M_{S_1}$.  
The region excluded by inverse-decay washout exhibits a sock-like shape, in which both inverse-decay processes are efficient; its lower edge is controlled by the $W_R^*$ branch becoming efficient while the upper feature occurs when the inverse-decay rates become suppressed by the Boltzmann factor. The kink of the boundary around $v_R \sim 10^5$\,GeV occurs when the required reheating temperature crosses $T_{\rm EW}$, so that $T_{\rm test}=\max(T_{\rm RH},T_{\rm EW})$ switches between $T_{\rm EW}$ and $T_{\rm RH}$.
The doubly off-shell scattering washout appears in the low-$v_R$ region.

\begin{figure}[t!]
    \centering
    \includegraphics[width=0.8\linewidth]{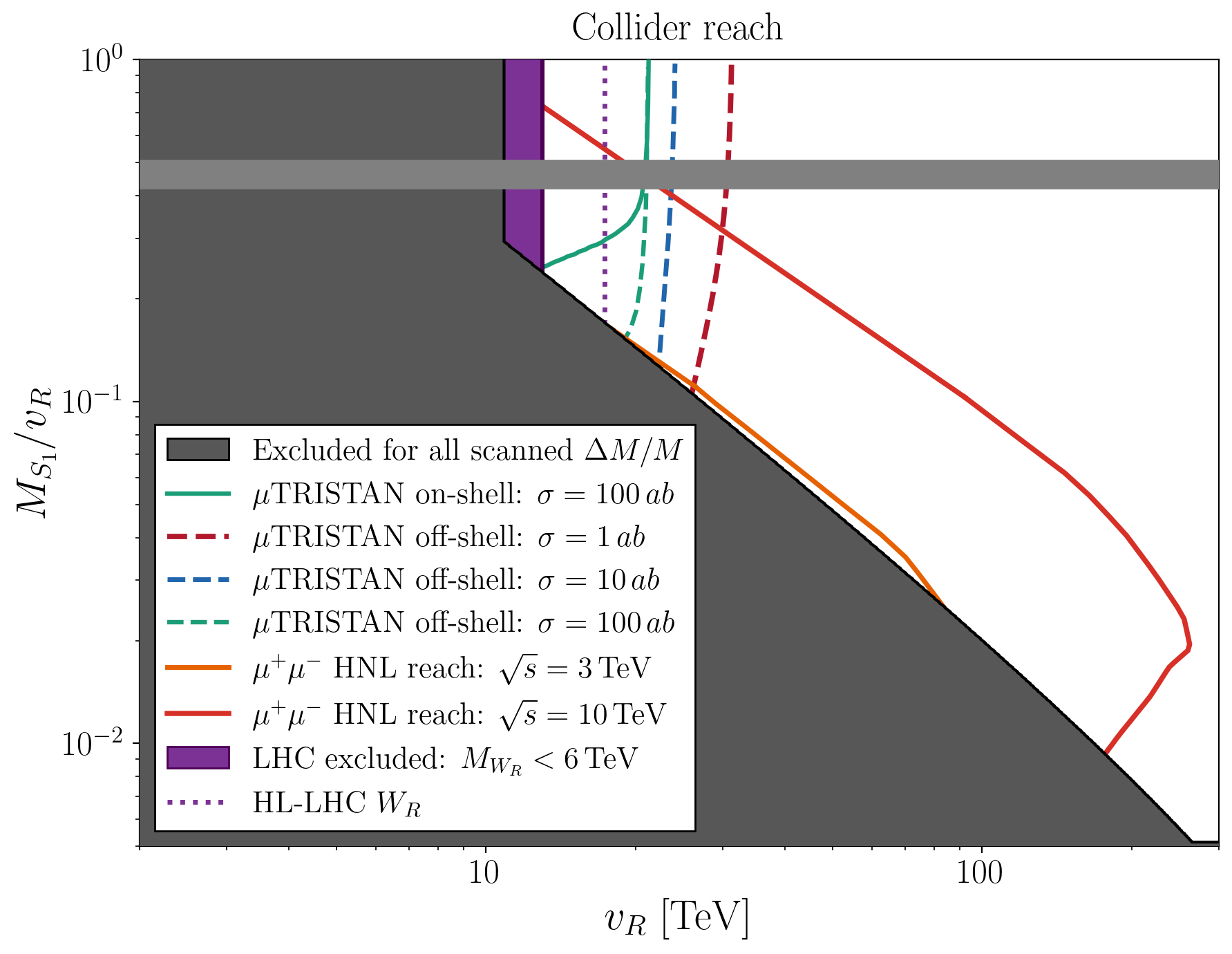}
    \caption{Collider reach overlaid on the non-thermal leptogenesis parameter space. The dark gray region is excluded for all scanned mass splittings, while the white region remains viable for at least one. The $\mu$TRISTAN contours show the projected $W_R$ sensitivity from same-sign $\mu^+\mu^+$ collisions, while the solid orange and red curves show the projected heavy neutral lepton (HNL) reach at $3\,$ TeV and $10\,$TeV opposite-sign $\mu^+\mu^-$ colliders, respectively, for the benchmark $|U_{\mu1}|^2=(v_L/v_R)^2$. The purple shaded region denotes the current LHC exclusion and the dotted purple line the HL-LHC reference reach. The horizontal gray band marks the on-shell/off-shell $W_R$ crossover.}
    \label{fig:collider}
\end{figure}

The surviving non-thermal region is directly comparable to collider probes of new gauge bosons and right-handed neutrinos. In figure~\ref{fig:collider}, we overlay the cosmologically viable region onto the current constraints and projected sensitivity of collider experiments.
The current LHC lower bound on $M_{W_R} > 6\,$TeV from $W_R \to  e \nu$ excludes the lowest $v_R$ corner of parameter space, while the HL-LHC is projected to be sensitive to $M_{W_R} \sim 8\,$TeV~\cite{ATLAS:2019lsy,ATLAS:2023ibb,CMS:2018jxx,CMS:2021ctt,CidVidal:2018eel}. A part of the remaining viable non-thermal parameter space can be probed by the $\mu$TRISTAN experiment~\cite{Hamada:2022mua}, especially for $v_R$ in the few-tens-of-TeV range.
The projected $\mu$TRISTAN contours are taken from the  lepton-number-violating process $\mu^+ \mu^+ \to W_R^{+(*)} W_L^+$ at $\sqrt{s}=10\,\mathrm{TeV}$, with on-shell and off-shell regimes shown separately~\cite{Harigaya:2025zru}.
Opposite-sign $\mu^+\mu^-$ colliders provide a complementary probe of $S_1$ through the mixing of it with active neutrinos. For $\lambda^S v_R \gg M^{\rm Maj}$, the characteristic active-heavy mixing scale is set by $v_L/v_R$~\cite{Hall:2023vjb}. We overlay the projected $3\,$TeV and $10\,$TeV $\mu^+\mu^-$ collider sensitivity to a muon-flavored heavy neutral lepton from Ref.~\cite{Li:2023tbx}, identifying $m_N=M_{S_1}$ and adopting the unsuppressed muon-flavor benchmark $|U_{\mu1}|^2=(v_L/v_R)^2$.

We note that there may be extra contributions to baryon asymmetry, depending on the decay mode of the inflaton. For example, if the inflaton is a real scalar and has couplings $ y^\phi_{11}\phi S_1 S_1$ and $y^\phi_{12}\phi S_1 S_2$, quantum corrections to the decay $\phi \rightarrow S_1 S_1$ by $y^\phi_{12}$ and $y^S_{12}$ give $\Gamma(\phi \rightarrow S_1 S_1) \neq \Gamma(\phi \rightarrow S_1^\dagger S_1^\dagger)$. $S_1$ then dominantly decays into $\ell^\dagger H_L^\dagger$ to produce lepton asymmetry. Here the lepton number is violated by the coexistence of $y^\phi$ and $y^S$. Since the CP asymmetry is not suppressed by the smallness of the branching ratio into the $W_R$ mode, the viable $v_R$ for $M \ll v_R$ can be smaller  than that in figure~\ref{fig:nontherm}. For $M\sim v_R$, which minimizes the viable $v_R$ as can be seen in figure~\ref{fig:nontherm}, the lepton asymmetry produced by CP-violating decay of $S$ is not suppressed, so the lowest possible $v_R$ cannot be lowered with the extra production of asymmetry by the inflaton.

\FloatBarrier

\section{\label{sec:fin}Conclusions}

In this work, we studied leptogenesis in the minimal Higgs parity solution to the strong CP problem. In this framework, neutrino masses are radiatively generated through the interplay of right-handed neutrinos and their gauge-singlet Dirac partners. After $SU(2)_R$ breaking, the right-handed neutrinos decay via two channels: their Yukawa interactions and $W_R$ exchange. The coexistence of these two decay modes violates lepton number, while CP symmetry is violated by the Yukawa couplings --- the two essential ingredients for leptogenesis.

The viable parameter space, spanned by the $SU(2)_R$ breaking scale $v_R$ and the lightest right-handed neutrino mass $M_{S_1}$, depends on the cosmological origin of the right-handed neutrino abundance. For thermal production, the requirement of sufficient departure from equilibrium forces $v_R \gtrsim 6\times 10^{12}$\,GeV, far beyond the reach of current or proposed $W_R$ and right-handed neutrino searches. This is because the same interactions responsible for generating a sizable CP asymmetry also maintain the right-handed neutrinos in thermal equilibrium, suppressing the out-of-equilibrium condition. The situation is further constrained at low $M_{S_1}/v_R$, where the off-shell $W_R^*$-mediated decay is suppressed while the $H_L$ channel remains efficient, reducing the net lepton-number violation. Nevertheless, the large $v_R$ required in the thermal case is consistent with the Higgs Parity mechanism's prediction, in which the running SM Higgs quartic coupling is predicted to vanish near the scale $v_R$.

For non-thermal production via inflaton decay, the right-handed neutrino abundance is no longer tied to thermal equilibrium. The Universe's temperature at the time of decay can be well below $M$, allowing the decays to proceed out of equilibrium even when the CP-violating Yukawa couplings are sizable. As a result, successful leptogenesis remains viable with $SU(2)_R$ breaking at the multi-TeV scale, even without relying on resonant enhancement of CP asymmetry. The lowest cosmologically viable values of $v_R$ overlap with existing LHC constraints on the $W_R$ mass, while a surviving region in the tens-of-TeV range remains accessible to future experiments, including the HL-LHC and the same-sign $\mu^+\mu^+$ collider  proposal ($\mu$TRISTAN). Opposite-sign $\mu^+\mu^-$ colliders provide a complementary probe through direct searches for the heavy neutral leptons. This model therefore provides a concrete example in which parity symmetry and radiative neutrino masses combine to connect the origin of the baryon asymmetry to experimentally accessible right-handed gauge bosons and heavy neutral leptons.

The viability of leptogenesis in this model ultimately rests on the structure of the charged and neutral lepton masses in the minimal Higgs framework. Because the radiative origin of neutrino masses decouples the Yukawa couplings from the light-neutrino mass scale, the CP- and lepton-number-violating couplings can be large even for small right-handed neutrino masses. Furthermore, while the CP-violating decay would be absent at one loop if the right-handed charged leptons originated exclusively from the $SU(2)_R$ partners of the right-handed neutrinos, the minimal Higgs model generically produces right-handed charged leptons as mixtures of $SU(2)_R$ partners and states in other $SU(2)_R$ representations. This mixing allows CP violation to arise at one loop level, which is central to the mechanism.

Several refinements remain for future work. A full flavor-dependent Boltzmann treatment and dedicated finite-temperature calculations could shift the present boundaries by $\mathcal{O}(1)$ factors, though the central conclusions are not expected to change.

The low-scale $SU(2)_R$ breaking opens further connections between parity symmetry and beyond-the-Standard-Model phenomenology. In particular, the non-thermal $v_R$ window is consistent with the upper bound on $v_R$ in WIMP models where dark matter annihilates via new gauge-boson exchange~\cite{Baldwin:2025oqt,Baldwin:2026mwv}. The light right-handed neutrinos and $W_R$ boson can mediate neutrinoless double beta decay~\cite{Blennow:2010th,Barry:2013xxa,Harigaya:2025zru}, and the large Yukawa couplings of the light right-handed neutrinos can induce observable rates for $\mu \to e\gamma$. The detailed predictions for these rare processes depend on the lepton flavor structure and are left for future work.

\begin{acknowledgments}
KH was supported by the Department of Energy grant DE-SC0009924 and World Premier International Research Center Initiative (WPI), MEXT, Japan (Kavli IPMU). GL was supported by a generous fellowship from the University of Chicago Department of Physics for a portion of this work.
\end{acknowledgments}

\appendix

\section{\label{sec:AppendixCP} CP-violation parameter}
In our calculation of the CP-violation parameter, we consider the self-energy contributions to $\epsilon_{ia}$, including the possibility of two nearly degenerate singlet masses $M_{S_i}\approx M_{S_k}$. We use the resummation methods outlined in \cite{Pilaftsis:1997jf,Pilaftsis:1998pd,Pilaftsis:2003gt} to properly treat the absorptive part of the $S_i$ decays with self-energy corrections. We begin from the radiative singlet interaction Lagrangian in Eq.~\eqref{RadL}. 

We define the CP-violation parameter per channel as in Eq.~\eqref{eq:channelCP}:
\begin{equation}
\label{eq:channelCP_appA}
\epsilon_{ia}\equiv
\frac{\Gamma(S_i \to \bar{e}_a^\dagger W_R^\dagger) - \Gamma\left(S^\dagger_i \to \bar{e}_a W_R\right)}
{\Gamma(S_i \to \bar{e}_a^\dagger W_R^\dagger) + \Gamma \left(S^\dagger_i \to \bar{e}_a W_R\right)}.
\end{equation}
At tree level, the decay $S_i \to \bar{e}_a^\dagger W_R^\dagger$ is mediated by the right-handed gauge interaction together with the mixing in the $\bar e$ sector. It is convenient to write the tree-level transition element as
\begin{equation}\label{eq:Ttree_appA}
\mathcal{T}^{(0)}_{ia}
=
\frac{g_{2,R}}{\sqrt{2}}
(\tilde U^{\bar{e}}_{ia})\,
\varepsilon^{\mu\,*} (p_W)\,
\bar{u}(p_e)\gamma_\mu P_R u(p_S),
\end{equation}
Here $\gamma_\mu$ are Dirac matrices, $P_{R,L}=(1\pm\gamma_5)/2$, $\varepsilon^\mu(p_W)$ is the $W_R$ polarization vector, and $u,\bar u$ are Dirac spinors with $\bar u=u^\dagger\gamma^0$. With the definition \eqref{eq:channelCP_appA} and $\Gamma^{(0)}$ being the corresponding tree-level decay width of $S_i \to \bar{e}_a^\dagger W_R^\dagger$, the denominator is well-approximated by $2\Gamma^{(0)}$ to leading nontrivial order in the loop expansion, so that $\epsilon_{ia}$ is controlled by the interference between the tree and one-loop amplitudes.

We now consider the one-loop self-energy diagram shown in figure~\ref{fig:Feynman}, which induces mixing $S_i\to S_k$ through the $\ell_bH_L$ cut. We denote the corresponding absorptive (cut) part of the transition as $\Sigma^{H_L,\,\mathrm{abs}}_{ki}$. The standard Cutkosky result gives \cite{Cutkosky:1960sp,Pilaftsis:1997jf}
\begin{equation}\label{eq:ImSigma_appA}
    \Sigma^{H_L,\,\mathrm{abs}}_{ki}(\slashed p) = \frac{1}{16\pi}\, \left(y^S y^{S\dagger}\right)_{ki}\, \slashed p\,P_L, \qquad \left(y^S y^{S\dagger}\right)_{ki} = \sum_b y^S_{kb}\,y^{S*}_{ib}.
\end{equation}
This is where the complex phases in $y^S$ enter the self-energy diagram at leading order. In particular, the factor $1/(16\pi)$ belongs to the absorptive self-energy insertion itself, in direct analogy with the absorptive mixing coefficient used in the resummed treatment of unstable-particle mixing \cite{Pilaftsis:1997jf,Pilaftsis:1998pd,Pilaftsis:2003gt}. 

The intermediate $S_k$ propagator must be dressed by its diagonal absorptive self-energy. We denote the physical total width of $S_k$ by $\Gamma_k^{\mathrm{Tot}}$, which includes all open decay channels. In the resonant regime $M_{S_i}\approx M_{S_k}$, the propagator must be treated in the usual Breit--Wigner form, obtained by Dyson resummation of the diagonal self-energy \cite{Pilaftsis:1997jf,Pilaftsis:1998pd,Pilaftsis:2003gt,Dev:2017wwc,Garny:2011hg,Garbrecht:2011aw},
\begin{equation}\label{eq:BW_appA}
\frac{i}{\slashed p - M_{S_k}} \;\longrightarrow\; i\,\Pi_{S_k}(\slashed p) \simeq i\,\frac{\slashed p + M_{S_k}}{p^2 - M_{S_k}^2 + i\,M_{S_k}\Gamma_k^{\mathrm{Tot}}}.
\end{equation}
This regulates the would-be divergence as $M_{S_i}\to M_{S_k}$. The self-energy contribution to the decay amplitude is obtained by inserting the mixing transition \eqref{eq:ImSigma_appA} and the resummed propagator \eqref{eq:BW_appA} into the tree-level decay of $S_k\to \bar e_a^\dagger W_R^\dagger$. At the amplitude level,
\begin{equation}\label{eq:Tloop_appA}
    \mathcal{T}^{(1)}_{ia} = \frac{g_{2,R}}{\sqrt{2}}\, \varepsilon^{\mu\,*} (p_W)\, \bar{u}(p_e)\gamma_\mu P_R \left[ \sum_{k\neq i} (\tilde U^{\bar e}_{ka})\, \Pi_{S_k}(\slashed p_S)\, \Sigma^{H_L,\,\mathrm{abs}}_{ki}(\slashed p_S) \right] u(p_S),
\end{equation}
where the sum is over the intermediate singlet index $k$. The CP-conjugate amplitude is obtained by complex conjugating the couplings and replacing the external spinors by their charge-conjugate counterparts. After summing over spins, the corresponding kinematic factors are identical.

To leading nontrivial order, the difference between the decay width and its CP conjugate is controlled by the interference of $\mathcal{T}^{(0)}_{ia}$ and $\mathcal{T}^{(1)}_{ia}$. Specifically, the interference contribution to the decay width contains $\mathcal T^{(0)}_{ia}\mathcal T^{(1)*}_{ia}$ and therefore the complex conjugate of the absorptive mixing insertion. In the ratio \eqref{eq:channelCP_appA}, the common spinor and phase-space factors cancel, and one obtains a result that can be written purely in terms of couplings and the resonant regulator from the dressed propagator \cite{Pilaftsis:1997jf,Pilaftsis:1998pd,Pilaftsis:2003gt,Dev:2017wwc,Garny:2011hg,Garbrecht:2011aw}. Keeping only the self-energy contribution shown in figure~\ref{fig:Feynman}, the channel asymmetry takes the resonant-safe form \cite{Pilaftsis:1997jf,Pilaftsis:1998pd,Pilaftsis:2003gt,Dev:2017wwc,Garny:2011hg,Garbrecht:2011aw}
\begin{equation}
\label{eq:resCPFirstOrder_appA}
    \epsilon_{ia} = \frac{1}{\left|\tilde U^{\bar e}_{ia}\right|^2} \sum_{\substack{b\\k\neq i}} \frac{1}{16 \pi}\, \operatorname{Im} \left( \tilde U^{\bar e}_{ia}\,{\tilde U^{\bar e\,{*}}_{ka}}\, y^{S}_{ib}\,y^{S\,*}_{kb} \right) \, \frac{M_{S_i}M_{S_k}\left(M_{S_i}^2 - M_{S_k}^2 \right)} {\left(M_{S_i}^2 - M_{S_k}^2 \right)^2 + \left(M_{S_k}\Gamma^{\mathrm{Tot}}_k\right)^2},
\end{equation}
which peaks when $|M_{S_i}^2-M_{S_k}^2|\sim M_{S_k}\Gamma_k^{\mathrm{Tot}}$. Note that we took the tree-level $H_L$-mediated decay to dominate the decay width. Furthermore, in the limit where $|M_{S_i}^2-M_{S_k}^2|\gg M_{S_k}\Gamma_k^{\mathrm{Tot}}$, Eq.~\eqref{eq:resCPFirstOrder_appA} reduces to the non-resonant scaling
\begin{equation}
\label{eq:nonres_appA}
    \epsilon_{ia} \to \frac{1}{\left|\tilde U^{\bar e}_{ia}\right|^2} \sum_{\substack{b\\k\neq i}} \frac{1}{16 \pi}\, \operatorname{Im} \left( \tilde U^{\bar e}_{ia}\,{\tilde U^{\bar e\,*}_{ka}}\,y^{S}_{ib} \, y^{S\,*}_{kb} \right) \, \frac{M_{S_i}M_{S_k}}{M_{S_i}^2 - M_{S_k}^2}.
\end{equation}

\section{\label{sec:AppendixBoltz}Boltzmann equations}
We consider the Boltzmann collision term governing the number density of species $X$: $\frac{d}{dt} n_{X} + 3 Hn_{X}= C\left[n_{X} \right]$, where $H$ is the temperature-dependent Hubble parameter.  We assume that the phase-space density $f_X$ of species $X$ is in the Maxwell--Boltzmann limit, with the equilibrium phase-space density $f^{\mathrm eq}_X$ satisfying $f_X/f^{eq}_X = \exp{\left(\mu_X/T\right)}\simeq 1+ \mu_X/T $, where $\mu_X$ is the chemical potential of species $X$ and $T$ is the temperature of the species. We define the dimensionless $z=M_{S_1}/T$ and yield $Y_{X} = n_{X}/s$, with $s$ being the entropy density. This gives the convenient relationship for the Boltzmann collision equation
\begin{equation}
    s\frac{d Y_X}{dt} = \frac{d n_X}{dt} + 3 Hn_X = C\left[n_{X} \right].
\end{equation}
The number density for the particle asymmetries is generally expressed as $n_{\Delta X}$ and is related to the corresponding chemical potential as $n_{\Delta X} = g_X (\mu_X /T)\, T^3/6$ for fermions and $n_{\Delta X} = g_X (\mu_X /T)\, T^3/3$ for bosons, where $g_X$ represents the internal degrees of freedom of $X$.
In radiation domination with a constant light degree of freedom, $H \propto T^2$ and thus $H = H(M_{S_1})/z^2$. With $dz/ dt = H(M_{S_1})/z$, we rewrite our Boltzmann equations as differential equations for $Y_X(z)$. For $S_i$ with a decay width $\Gamma_i$, we  take the thermally averaged decay width to follow $\left \langle \Gamma_i \right \rangle = \mathcal{K}_1(M_{S_i} \,z /M_{S_1})/\mathcal{K}_2(M_{S_i} \,z /M_{S_1})\, \Gamma_i$, where $\mathcal{K}_i$ is the $i^\mathrm{th}$ modified Bessel function of the second kind.

\subsection{\label{sec:BoltzS}Boltzmann equations for \texorpdfstring{$S$}{S} and \texorpdfstring{$S^\dagger$}{Sdag}}
\label{App:SBoltz}
We begin by considering the Boltzmann collision term governing the number densities of the singlet fermions $S$ and $S^\dagger$. For $M_{S_i}>M_{W_R}$, the individual collision terms are
\begin{equation}
    \begin{aligned}
        C[n_{S_i}] & = -n_{S_i} \left(\left\langle \Gamma^{H_L}_{ia}\right\rangle 
        +\left\langle \Gamma^{W_R}_{ia}\right\rangle\right) + n_{S_i}^{\mathrm eq} \left(\frac{n_{\ell_a}n_{H_L}}{n_{\ell_a}^{\mathrm eq}n_{H_L}^{\mathrm eq}}\left\langle \Gamma^{H_L^\dagger}_{ia}\right\rangle + \frac{n_{\bar{e}_a}}{n_{\bar{e}_a}^{\mathrm eq}}\left\langle \Gamma^{W^\dagger_R}_{ia}\right\rangle \right),\\
        C[n_{S^\dagger_i}] & = -n_{S^\dagger_i} \left(\left\langle \Gamma^{H_L^\dagger}_{ia}\right\rangle 
        +\left\langle \Gamma^{W^\dagger_R}_{ia}\right\rangle\right) + n_{S_i}^{\mathrm eq}\left(\frac{n_{\ell^\dagger_a}n_{H_L^\dagger}}{n_{\ell_a}^{\mathrm eq}n_{H_L}^{\mathrm eq}}\left\langle \Gamma^{H_L}_{ia}\right\rangle + \frac{n_{\bar{e}^\dagger_a}}{n_{\bar{e}_a}^{\mathrm eq}}\left\langle \Gamma^{W_R}_{ia}\right\rangle \right),
    \end{aligned}
\end{equation}
where we implicitly sum over all final states $a$ and set the chemical potential of all gauge bosons to be zero. It is convenient to define the CP-even and CP-odd number densities $n_{S_+} $ and $n_{S_-}$ as $n_{S_\pm} =\left(n_{S} \pm  n_{S^\dagger})\right/2$. It follows that

\begin{equation}
    \begin{aligned}
        C[n_{S_i^+}] & = -\left(n_{S_i^+} -  n_{S_i}^{\mathrm eq}\right)\left(\left\langle \Gamma^{H_L}_{ia}\right\rangle 
        +\left\langle \Gamma^{W_R}_{ia}\right\rangle\right),\\
        C[n_{S_i^-}] & = -n_{S_i^-} \left(\left\langle \Gamma^{H_L}_{ia}\right\rangle 
        +\left\langle \Gamma^{W_R}_{ia}\right\rangle\right) + n_{S_i}^{\mathrm eq} \left[\left(6\frac{n_{\Delta {\ell}_a}}{g_\ell T^3} + 3\frac{n_{\Delta H_L}}{g_H T^3} \right) \left\langle \Gamma^{H_L}_{ia}\right\rangle + 6\frac{n_{\Delta {\bar{e}}_a}}{g_{\bar{e}} T^3} \left\langle \Gamma^{W_R}_{ia}\right\rangle \right],
    \end{aligned}
\end{equation}
where we neglect higher-order terms in $\epsilon$ and $\mu/T$. The listed decay widths are computed at tree level, allowing us to use the simplification $\Gamma^X +\Gamma^{X^\dagger} = 2 \Gamma^X$. Converting into the dimensionless yield form, the relevant Boltzmann equations to leading order in CP violation are

\begin{equation}\label{eq:BoltzSmu}
\begin{aligned}
  \frac{dY_{S_i^+}}{dz}
  &= -\frac{z}{H(M_{S_1})}\sum_{a} \left(\left\langle\Gamma^{W_R}_{ia} \right\rangle+\left\langle\Gamma^{H_L}_{ia}\right\rangle\right) \left(Y_{S_i^+}-Y_{S_i}^{\mathrm{eq}}\right),\\
\frac{dY_{S_i^-}}{dz}
  &= -\frac{z}{H(M_{S_1})}\sum_{a}\Biggr\{\left(\left\langle\Gamma^{W_R}_{ia} \right\rangle+\left\langle\Gamma^{H_L}_{ia}\right\rangle\right)  Y_{S_i^-}\\
  &\qquad
  - 3 \frac{n_{S_i}^{\mathrm eq}}{T^3} \left[\left\langle \Gamma^{H_L}_{ia}\right\rangle \left(Y_{\Delta {\ell}_a}+ \frac{1}{2}Y_{\Delta H_L}\right) + 2 \left\langle \Gamma^{W_R}_{ia}\right\rangle Y_{\Delta {\bar{e}}_a}  \right]
  \Biggr\}.
\end{aligned}
\end{equation}
Since we assume $S_i$ follows Maxwell--Boltzmann statistics, the equilibrium number density is
\[
n_{S_i}^{\rm eq}(T)=g_{S_i}\,\frac{M_{S_i}^2 T}{2\pi^2}\,\mathcal K_2\!\left(\frac{M_{S_i}}{T}\right),
\]
and therefore, in terms of $z\equiv M_{S_1}/T$ and $x_i\equiv M_{S_i}/M_{S_1}$,
\[
\frac{n_{S_i}^{\rm eq}}{T^3}=\frac{g_{S_i}}{2\pi^2}\,(x_i z)^2\,\mathcal K_2(x_i z).
\] For $M_{S_i} \leq M_{W_R}$ where $S$ cannot decay into an on-shell $W_R$, one can replace $\left\langle\Gamma^{W_R}_{ia}\right\rangle$ with the off-shell mediated decay $\sum_{kb}\left\langle\Gamma^{\mathrm{Lep}}_{iakb}\right\rangle + \left\langle\Gamma^{\mathrm{Had}}_{ia}\right\rangle$.

\subsection{\label{sec:BoltzL}Boltzmann equations for \texorpdfstring{$\ell$}{l}, \texorpdfstring{$H_L$}{PhiL}, and \texorpdfstring{$\bar{e}$}{ebar}}

We begin with the case where $M_{S_i}>M_{W_R}$, separately keeping track of the net $\ell$, $H_L$, and $\bar{e}$ yields. We first consider the relevant collision term for the $n_{\Delta\bar{e}}$ contribution from the decays of $S$ and $S^\dagger$:

\begin{equation}\label{eq:decayCollisionEbar}
\begin{aligned}
    C\left[n_{\Delta\bar{e}_a} \right]_{\rm decay}
    &=  n_{S_i} \left\langle \Gamma^{W_R}_{ia}\right\rangle -n_{S^\dagger_i} \left\langle \Gamma^{W^\dagger_R}_{ia}\right\rangle\\
    &=2 \left\langle \Gamma^{W_R}_{ia}\right\rangle \left(\epsilon_{ia} n_{S_i^+}+ n_{S_i^-} \right),
\end{aligned}
\end{equation} where we implicitly sum over the initial state $i$ and compute the relevant decay width at tree level. The collision term for the $n_{\Delta \ell,\Delta H_L}$ contribution from the decays of $S$ and $S^\dagger$:
\begin{equation}\label{eq:decayCollisionl}
    \begin{aligned}
        C\left[n_{\Delta \ell_a, \Delta {H_L}} \right]_{\rm decay}
        &=  n_{S_i} \left\langle \Gamma^{H_L}_{ia}\right\rangle - n_{S^\dagger_i} \left\langle \Gamma^{H_L^\dagger}_{ia}\right\rangle\\
        &=-2 \left\langle \Gamma^{W_R}_{ia}\right\rangle \epsilon_{ia} n_{S_i^+}
        +2 \left\langle \Gamma^{H_L}_{ia}\right\rangle n_{S_i^-} 
    \end{aligned},
\end{equation}
where we additionally sum over the generational index $a$ for the $\Delta H_L$ collision term. For the contribution from the inverse decays of $\ell \,H_L$, $\bar e\, W_R$, and their CP conjugates, one can perform a similar Boltzmann equation analysis and find
\begin{equation}\label{eq:washoutCollision}
    \begin{aligned}
       C\left[n_{\Delta\bar{e}_a} \right]_{\rm inverse}
        &= -2 \left\langle \Gamma^{W_R}_{ia}\right\rangle n_{S_i}^{\mathrm eq} \left(6\frac{n_{\Delta \bar{e}_a}}{g_{\bar{e}}T^3}  - \epsilon_{ia}\right),\\
       C\left[n_{\Delta \ell_a, \Delta {H_L}} \right]_{\rm inverse}
        &= -2 n_{S_i}^{\mathrm eq} \left[\epsilon_{ia} \left\langle \Gamma^{W_R}_{ia}\right\rangle + \left(6\frac{n_{\Delta {\ell}_a}}{g_\ell T^3} + 3\frac{n_{\Delta H_L}}{g_H T^3} \right) \left\langle \Gamma^{H_L}_{ia}\right\rangle \right],
    \end{aligned}
\end{equation}
having set the chemical potential of the gauge bosons to be zero. 

Combining Eq.~\eqref{eq:decayCollisionEbar}, Eq.~\eqref{eq:decayCollisionl}, and Eq.~\eqref{eq:washoutCollision}, the resultant collision term would incorrectly generate a lepton asymmetry even when the singlets are in thermal equilibrium. This is the usual real-intermediate-state double-counting problem. We therefore include the near-resonant contributions from the $\ell H_L \leftrightarrow \bar e W_R$ scatterings and their CP-conjugate processes, with the real intermediate singlet contribution subtracted \cite{KOLB1980224,Buchmuller:1997yu,Giudice:2003jh,Buchmuller:2004nz,Hahn-Woernle:2009jyb,Beneke:2010wd}. We can write the inverse-decay washout terms in a form that makes detailed balance manifest. The inverse decay $\bar e_a^\dagger W_R^\dagger \leftrightarrow S_i$
relaxes the chemical-potential difference $\mu_{S_i}-\mu_{\bar e_a}$, where we have taken $\mu_{W_R}=0$. Taking the Maxwell--Boltzmann approximation, the corresponding washout term appears in the combination $Y_{S_i^-} - 6n_{S_i}^{\rm eq}Y_{\Delta\bar e_a}/(g_{\bar e}T^3)$. Similarly, the inverse decay $\ell_\alpha H_L\leftrightarrow S_i$
relaxes $\mu_{S_i}-\mu_{\ell_\alpha}-\mu_{H_L}$. We find the corresponding washout combination to be $Y_{S_i^-} - n_{S_i}^{\rm eq} \left({6Y_{\Delta\ell_\alpha}}/{g_\ell}+{3Y_{\Delta H_L}}/{g_H} \right)/T^3$. Thus, to leading order in CP violation, the Boltzmann equations are
\begin{equation}\label{eq:BoltzDerived}
    \begin{aligned}
    \frac{d Y_{\Delta \bar{e}_a}}{dz}
    &= \frac{2z}{H(M_{S_1})} \sum_i \left\langle \Gamma^{W_R}_{ia}\right\rangle \Biggl[ \epsilon_{ia} \left(Y_{S_i^+}-Y^{\rm eq}_{S_i}\right) +Y_{S_i^-}
    \\
    &\hspace{4.2cm}
    - \frac{6n_{S_i}^{\rm eq}}{g_{\bar e}T^3} Y_{\Delta \bar{e}_a} \Biggr],
    \\
    \frac{d Y_{\Delta \ell_\alpha}}{dz}
    &=
    \frac{2z}{H(M_{S_1})} \sum_i \Biggl\{ -\epsilon_{i\alpha} \left\langle \Gamma^{W_R}_{i\alpha}\right\rangle \left(Y_{S_i^+}-Y^{\rm eq}_{S_i}\right)
    \\
    &\hspace{2.2cm}
    + \left\langle \Gamma^{H_L}_{i\alpha}\right\rangle \Biggl[ Y_{S_i^-}
    - n_{S_i}^{\rm eq}
    \left( \frac{6Y_{\Delta \ell_\alpha}}{g_\ell T^3} + \frac{3Y_{\Delta H_L}}{g_H T^3} \right) \Biggr] \Biggr\},
    \\
    \frac{d Y_{\Delta H_L}}{dz}
    &= \frac{2z}{H(M_{S_1})} \sum_{i,\alpha} \Biggl\{
    -\epsilon_{i\alpha} \left\langle \Gamma^{W_R}_{i\alpha}\right\rangle \left(Y_{S_i^+}-Y^{\rm eq}_{S_i}\right)
    \\
    &\hspace{2.2cm}
    + \left\langle \Gamma^{H_L}_{i\alpha}\right\rangle \Biggl[ Y_{S_i^-} - n_{S_i}^{\rm eq} \left( \frac{6Y_{\Delta \ell_\alpha}}{g_\ell T^3} + \frac{3Y_{\Delta H_L}}{g_H T^3} \right) \Biggr] \Biggr\}.
    \end{aligned}
\end{equation}
For our treatment, we take $g_{\bar e}=1$ and $g_\ell=g_H=2$. With this convention the $H_L$-mediated washout term reduces to $Y_{S_i^-} - 3{n_{S_i}^{\rm eq}}\left(Y_{\Delta\ell_\alpha}+\frac12 Y_{\Delta H_L}\right)/T^3$, which is the form used in the main numerical analysis.

In the case where $M_{S_i}\leq M_{W_R}$, the $W_R$-mediated decay proceeds through off-shell three-body channels. The leptonic channel $S_i\to \bar e_b^\dagger \bar\nu_k^\dagger \bar e_c$ changes the net $\bar e_a$ number by $\delta_{ab}-\delta_{ac}$, while the hadronic channel $S_i\to \bar e_b^\dagger \bar u\bar d^\dagger$ changes it by $\delta_{ab}$. We include the off-shell $W_R$-mediated widths in the CP-odd source terms, while neglecting their inverse-decay washout in the parameter region where the $H_L$-mediated inverse decays dominate. The resulting asymmetry equations are
\begin{equation}\label{eq:BoltzDerivedOffShell}
    \begin{aligned}
    \frac{d Y_{\Delta \bar{e}_a}}{dz}
    &= \frac{2z}{H(M_{S_1})}\sum_{i,b} \left[\epsilon_{ib} \left(Y_{S_i^+}-Y^{\rm eq}_{S_i}\right) + Y_{S_i^-} \right]
    \\
    &\hspace{1.4cm}
    \times \Biggl[ \sum_{k,c} \left(\delta_{ab}-\delta_{ac}\right) \left\langle\Gamma^{\rm Lep}_{ibkc}\right\rangle
    + \delta_{ab}
    \left\langle\Gamma^{\rm Had}_{ib}\right\rangle \Biggr],
    \\ \frac{d Y_{\Delta \ell_\alpha}}{dz}
    &= \frac{2z}{H(M_{S_1})}  \sum_i \Biggl\{ -\epsilon_{i\alpha} \left( \sum_{k,c} \left\langle\Gamma^{\rm Lep}_{i\alpha kc}\right\rangle + \left\langle\Gamma^{\rm Had}_{i\alpha}\right\rangle \right) \left(Y_{S_i^+}-Y^{\rm eq}_{S_i}\right)
    \\
    &\hspace{1.4cm}
    + \left\langle \Gamma^{H_L}_{i\alpha}\right\rangle \Biggl[ Y_{S_i^-}
    - n_{S_i}^{\rm eq} \left( \frac{6Y_{\Delta \ell_\alpha}}{g_\ell T^3} + \frac{3Y_{\Delta H_L}}{g_H T^3} \right) \Biggr] \Biggr\},
    \\ \frac{d Y_{\Delta H_L}}{dz}
    &= \frac{2z}{H(M_{S_1})} \sum_{i,\alpha} \Biggl\{ -\epsilon_{i\alpha} \left( \sum_{k,c} \left\langle\Gamma^{\rm Lep}_{i\alpha kc}\right\rangle + \left\langle\Gamma^{\rm Had}_{i\alpha}\right\rangle \right) \left(Y_{S_i^+}-Y^{\rm eq}_{S_i}\right)
    \\
    &\hspace{1.4cm}
    + \left\langle \Gamma^{H_L}_{i\alpha}\right\rangle \Biggl[ Y_{S_i^-}
    - n_{S_i}^{\rm eq} \left( \frac{6Y_{\Delta \ell_\alpha}}{g_\ell T^3}  + \frac{3Y_{\Delta H_L}}{g_H T^3} \right) \Biggr] \Biggr\},
    \end{aligned}
\end{equation}
where $\delta$ is the Kronecker delta.

\bibliographystyle{JHEP}
\bibliography{biblio.bib}% Produces the bibliography via BibTeX.

\end{document}